# The InSight-HP³ Mole on Mars: Lessons Learned from Attempts to Penetrate to Depth in the Martian Soil.

Tilman Spohn, International Space Science Institute, Bern, Switzerland and DLR Institute of Planetary Research, Berlin, Germany.

Troy L. Hudson, Jet Propulsion Laboratory, California Institute of Technology, Pasadena, USA

Lars Witte, DLR Institute of Space Systems, Bremen, Germany

Torben Wippermann, DLR Institute of Space Systems, Bremen, Germany

Lukasz Wisniewski, Astronika, Warsaw, Poland

Bartosz Kedziora, Astronika, Warsaw, Poland

Christos Vrettos, Department of Civil Engineering, University of Kaiserslautern, Kaiserslautern, Germany.

Ralph D. Lorenz, Johns Hopkins Applied Physics Laboratory, Laurel, MD 20723, USA

Matthew Golombek, Jet Propulsion Laboratory, California Institute of Technology, Pasadena, USA

Roy Lichtenheldt, DLR Institute of System Dynamics and Control, Oberpfaffenhofen, Germany

Matthias Grott, DLR Institute of Planetary Research, Berlin, Germany.

Jörg Knollenberg, DLR Institute of Planetary Research, Berlin, Germany.

Christian Krause, DLR Microgravity User Support Center, Cologne, Germany.

Cinzia Fantinati, DLR Microgravity User Support Center, Cologne, Germany.

Seiichi Nagihara, Department of Geosciences, Texas Tech University, Lubbock, USA.

Jurek Grygorczuk, Astronika, Warsaw, Poland.

## Abstract

The NASA InSight lander mission to Mars payload includes the Heat Flow and Physical Properties Package HP³ to measure the surface heat flow. The package was designed use a small penetrator - nicknamed the mole - to implement a vertical string of temperature sensors in the soil to a depth of 5 m. The mole itself is equipped with sensors to measure a thermal conductivity- depth profile as it proceeds to depth. The heat flow would be calculated from the product of the temperature gradient and the thermal conductivity. To avoid the perturbation caused by annual surface temperature variations, the measurements would be taken at a depth between 3 m and 5 m. The mole had been designed to penetrate cohesionless soil similar in rheology to Quartz sand which was expected to provide a good analogue material for Martian sand. The sand would provide friction to the buried mole hull to balance the remaining recoil of the mole hammer mechanism that drives the mole forward. Unfortunately, the mole did not penetrate more than roughly a mole length of 40 cm. The failure to penetrate deeper was largely due to a cohesive duricrust of a few tens of centimeter thickness that failed to provide the required friction. Although a suppressor mass and spring as part of the mole hammer mechanism absorbed much of the recoil, the available mass did not allow designing a system that would have eliminated the recoil. The mole penetrated to 40 cm depth benefiting from friction provided by springs in the support structure from which it was deployed. It was found in addition that the Martian soil provided unexpected levels of penetration resistance that would have motivated to designing a more powerful mole. The low weight of the mole support structure was not enough to guide the mole penetrating vertically. It is concluded that more mass would have allowed to design a more robust system with little or no recoil, more energy of the mole hammer mechanism and a more massive support structure. In addition, to cope with duricrust a mechanism to support the mole to a depth of about two mole lengths should be considered.

1. Introduction and HP³ Science Goals

The NASA Mars mission Interior Exploration using Seismic Investigations, Geodesy and Heat Transport (InSight) payload includes the Heat Flow and Physical Properties Package, HP³ (see Spohn et al., 2018 for a description of the package and its science goals). The Level 1 science goal of the package, and a Level 1 science goal of the mission - was to measure the surface heat flow out of the interior of Mars. The surface heat flow is a fundamental geophysical quantity constraining the level of geodynamic activity of a planet, its evolution and the energy balance of its interior and even its composition (e.g., Plesa et al., 2016, 2018).

The surface heat flow $F_s$ can be determined by measuring the near surface temperature gradient $dT/dz$, where $T$ is temperature and $z$ is depth, and the thermal conductivity $k$:

1) $$F_s = k \frac{dT}{dz}$$

The average surface heat flow on Mars has been estimated to be not much more than $25\ mW/m^2$ (e.g., Grott et al., 2012; Plesa et al., 2016, 2018; Smrekar et al., 2018). For a thermal conductivity of the order of $10^{-2}\ W/mK$, the expected temperature gradient will then be not more than a few Kelvin per meter.

Daily and annual surface temperature variations cause disturbances of the long-term temperature gradient at shallow depth. Therefore, the temperature gradient has to be measured at a depth where the disturbance is small enough to allow for the targeted measurement accuracy. For HP³, with $\pm 5\ mW/m^2$ as the targeted accuracy, the minimum tip depth was estimated to be $3\ m$ (Spohn et al., 2018). Practical considerations of mass and volume as well as planetary protection rules limited the depth to $5\ m$. The MEPAG Special Regions Science Analysis Group (Rummel et al., 2014) found that the depth to buried ice in the tropics and mid-latitudes on Mars would be > 5m.

To get the sensors to their target depths, which require a mole tip depth between 3–5 m, HP³ would use a small penetrator of 40 cm length and 2.7 cm diameter, nicknamed the mole. The penetrator was to pull a flat cable equipped with temperature sensors into the subsurface to measure the temperature gradient. The mole was further furnished with resistance-foil temperature sensors integrated in its hull that could be actively heated at a known constant power. By measuring the temperature increase over time, the thermal conductivity could be calculated from the data (Spohn et al., 2018; Grott et al., 2019, 2021). With the sensors on the mole and the tether it was planned to measure a thermal conductivity profile up to *5* m depth with a 50 cm resolution as well as a temperature profile and its variation over a Martian year. The package was supplemented with a radiometer to measure the surface brightness temperature to supplement the data (e.g., Spohn et al., 2018; Piqueux et al., 2021, Müller et al., 2021).

The mole technology is particularly suited for heat flow measurements with low to moderate cost. In comparison with rotational drilling, the mole uses much fewer resources of mass and volume on the spacecraft since a stabilizing rig is not required. Furthermore, the energy dissipated during penetration is significantly smaller for a mole than for a drill. This is important for the heat flow measurement since any energy dissipated upon penetration will disturb the temperature profile. There is no other way of removing the disturbance than to simply wait. This would be prohibitive for the measurement if the energy input and low thermal conductivity result in a dissipation time that is too long relative to the time constraints of the mission.

Another potentially useful technology for heat flow measurement is the pneumatic drill, which blows regolith particles out of the hole using gas jets, (Nagihara et al., 2021) under development for the Commercial Lunar Payload Services Program of NASA. Some lessons learned from the HP³ mole could be useful for this concept as well, but a detailed discussion is beyond the scope of the present paper.

InSight landed on the surface of Mars on November 26, 2018, and the package, initially mounted to the lander deck, was deployed to the surface on February 12, 2019 UTC (Sol 76 of the mission) using the InSight lander's robotic arm. The mole was released from its housing, the Support Structure (SS), in mid-February and began penetrating on March 1, 2019 UTC (Sol 92). For the first round of hammering, the mole was commanded to hammer to a tip depth of 70 cm or for a period of 4 hours, whichever came first.  It was believed, based on terrestrial tests, that the mole tip would reach the target depth of 70 cm after only few hundred hammer strokes (at 3.7 seconds per stroke, 30 minutes of hammering corresponds to ~486 strokes). However, the data returned from Mars suggested that the mole had not reached its target depth although 3881 hammer strokes had been successfully performed. The team, assuming that the mole had perhaps encountered a near surface layer of gravel or a stone, commanded a second round of hammering of 5 hours (4720 strokes). These were performed successfully but without any indication of progress. Instead, data taken by the STATIL tilt sensors inside the mole indicated that it had moved back and forth laterally (see Spohn et al. 2021 for details).

The failure of the mole to penetrate prompted the team to create an Anomaly Response Team (ART) consisting of JPL and DLR scientists and engineers.  In their attempts to understand the anomaly and assist the mole in penetrating further, the team used the InSight robotic arm to image, expose (through a lift and replacement of the mole Support Structure), and interact with the mole, as well as with the Martian soil.  The ART effort ended on January 9$^{th}$, 2021 (Sol 754), when a final test to see if the mole could penetrate without assistance from the arm failed. A more detailed account of the activities is given in the appendix.

The ART efforts were successful in getting the entire body of the mole below ground, where it will be used during the remainder of the mission as a near-surface thermal probe. This task of measuring the near-surface thermal conductivity of the Martian regolith over time is aimed at understanding how it evolves in response to changes in the ambient atmospheric pressure and temperature.  First results of this thermal conductivity experiment have been published in Grott et al. (2021). The almost 2 years of activities using the scoop of the InSight robotic arm to interact with the mole and the regolith, as well as coordinated observations with the seismometer instrument SEIS, have resulted in data on the properties of the Martian regolith at the InSight landing site (e.g., Golombek et al., 2018). These data are being described in a separate paper (Spohn et al., 2021).

In the present paper, we will discuss what lessons can be learned about the design and the operation of the mole from InSight. It is hoped that these can inform future attempts to use small penetrators, on Mars or other extraterrestrial bodies, whether for heat flow or other scientific and exploration purposes.

## 2. Development Time and Resources, Requirements and Expected Properties of the Martian Soil.

### 2.1 Development Time and Resources

The development of the HP$^3$ instrument had to occur under strong limitations in time and resources. Upon the selection of the InSight mission, it was thought that the instrument had a comfortable TRL of 5-6. A functional brassboard model developed in a pre-selection assessment phase had been successfully tested. This mole was termed the "Breadboard Mole" in the team's mole model terminology although it was a fully functional penetrator (i.e., more than a benchtop breadboard). The Breadboard Mole was based on the design of the PLUTO mole (e.g., Richter et al. 2002) developed for the ill-fated BEAGLE lander (part of the Mars Express mission), and on an industrial study done a few years before the InSight assessment phase. To add margin to the penetration performance it was decided to increase the stroke energy of the hammer mechanism. Unfortunately, it turned out that the Breadboard Mole design was not robust enough for the increased hammer strength, suffering many prototype failures when the mechanism's output energy was increased. This led to a major redesign of the hammer mechanism along with a reformation of the development team in late 2013, 2.5 years before the planned launch date in May 2016. The redesign was costly in terms of time and financial resources.

The limits of available funding and development time influenced decisions made during the development phase. For instance, when it became clear that penetration in sharp-edged high friction angle Syar[1] sand was difficult, it was decided that this regolith simulant would no longer be used because no data from Mars supported Syar and other high friction angle sands as representative of Martian sand. The sand was still used in some tests as a limiting case. An in-depth redesign of the Mole and its support structure such that the Mole would penetrate reliably into materials with high self-cohesion and friction angles was believed to be beyond the available time and resources (this was, of course, without knowledge of the impending launch delay to May, 2018 due to delivery problems with the SEIS instrument). In addition, to save time the project changed to the protoflight approach e.g., https://analyst.gsfc.nasa.gov/ryan/MOLA/definit.html although a full life test of the mole hardware was done.

After the launch had to be delayed by two years, the additional time was used to further improve the robustness of the mole, in particular, of the electric wiring and electrical and mechanical connectors with regard to the high-shock environment created by the hammer mechanism.

2.2 Requirements

Among the requirements defined for the mission and its payload, the following ones were of particular interest for developing the HP$^3$ mole:

| ID | Requirement |
|---|---|
| L5-TM-1 | The Mole shall penetrate to a depth of at least 3 m into representative Mars simulants ("MMS < 2 mm" and "WF-34") |
| L5-TM-3 | The Mole shall be capable of executing no fewer than 20,000 strokes in a regolith with properties as in L5-TM-1 |
| L5-TM-4 | The Mole shall be capable of reaching 3 m depth under earth ambient conditions into representative Mars simulants ("MMS < 2 mm" and "WF-34") within 24 h of cumulative hammering |

Table 1. Requirements of particular interest for the design of the HP$^3$ mole. See Wippermann, et al. 2020 for further descriptions of the regolith simulants "MMS < 2 mm" and "WF_34".

The flight model that was delivered to fly to Mars did satisfy these requirements. In particular, the mole flight hardware proved to be reliable and performed 12,000 hammer strokes on Mars without indications of functional and performance degradation. Life tests on flight-equivalent units performed in regolith simulants on Earth showed >60,000 strokes with no apparent degradation in hammer force.

An additional important motivator for operational decisions was the desire (not a formal requirement) for the mole to reach its minimum target depth of >3 m within 200 sols following landing. This derived from a desire to penetrating faster than thermal perturbations due to the lander's shadow and dust-clearing albedo changes during landing (Grott 2009).

The project requirement for the mole to penetrate 3-5 m, became a landing site selection requirement to find a location with at least this thickness of poorly consolidated soil or regolith (Golombek et al., 2017). A substantial effort was put into determining the thickness of the impact comminuted layer at the western Elysium Planitia landing site and estimating the probability of success based on measurements of rock abundance and fragmentation theory (Warner et al., 2017, Golombek et al., 2017, 2018, 2020).

---

[1] "Syar Industries" is the trademarked name of a building material supply company in northern California. Crushed basalt from the Lake Herman Volcanics Group was procured by the InSight project from Syar Industries as a bulk regolith simulant. The term 'Syar' was a shorthand used by the team to distinguish this simulant from other basaltic materials used in testing."

2.3 Expected Properties of the Martian Soil

Among the major unknowns in the development of the mole were the mechanical and thermal properties of the regolith in which it was expected to penetrate and measure the thermal conductivity and temperature. From the principles of the functioning of the mole (Krömer, et al. 2019), it was clear that the sand would have to be plastically deformable and to be able to flow around the mole as it descended. The mole had to be strong enough (as measured by its stroke energy, E) to allow a suitable penetration rate of 0.25 mm/stroke on average. This rate is derived from requirement L5-TM-4 to reach 3 m tip depth in less than 24 cumulative hours of hammering and a stroke frequency of the flight model of 0.27 Hz. From an energy balance perspective, the (plastic) deformation work done per stroke per unit area of the penetrator is to be provided by the energy of the mole

2) $$\sigma_P \cdot R = \frac{\varepsilon E}{\pi r_m^2}$$

where $\sigma_P$ is the penetration resistance per unit area, R the penetration distance per stroke, $\varepsilon$ the efficiency of turning energy per stroke into (plastic) deformation work, and $r_m$ the radius of the mole. Equation (2) does not account for the effect of friction along the tether on the penetration rate. Wippermann et al. (2020) report that such friction caused a significant reduction of the penetration rate at a depth of more than about 3 m in the Deep Penetration Test bed at DLR Bremen for Qz sand but not for Syar sand.  As this paper discusses the mole penetration performance at small pressure and depth << 3 m, the effect of friction on the tether is not further considered here.

As we will discuss in section 3, the mole needs friction on its hull to balance the recoil generated by its hammer mechanism. For soils with rheological properties that can be described by an internal friction angle $\phi$ and a cohesion c (so-called 'c-$\phi$ soils', section 4), the active soil pressure, the lateral pressure on the wall of a cavity and thus on the mole hull, decreases with increasing c and $\phi$ and can be zero at shallow depth depending also on the density of the soil.  The penetration resistance, as we will discuss in section 4, will depend on the overburden pressure in the soil but also on the rheological properties of the soil.

Because no data were available on the penetration resistance of Martian sands, the latter had to be estimated based on terrestrial analog materials and Martian soil mechanical properties. In Figure 1 we present a compilation of soil mechanical properties of Mars simulant sands (density, grain size, friction angle and cohesion) in comparison with data on Martian sands compiled by Herkenhoff et al., (2008) and Golombek et al. (2008).  Estimates of the angle of internal friction and cohesion of soils on Mars have been derived from interactions with arms, scoops and wheels (images, motor currents and wheel parameters) and have been related to thermal inertia data (for density and cohesion), and relations linking density to other soil parameters using empirical relations for lunar regolith.

The data for the simulant sands are generally from data sheets that came with the sands. Qz sand and a 80/20 Syar sand/dust mix used in the Deep Penetration Tests have been measured with triaxial shear tests (Vrettos, 2012) at the University of Kaiserslautern, Department of Civil Engineering to confirm the specifications.  The data for Mojave Mars simulant (MMS sand and dust) are from Peters et al. (2008). The test materials have been selected on the basis of availability and spread in soil physics parameters and were thought to bracket the parameters of soils expected at the InSight landing site (see also Delage et al., 2017).

As shown in Figure 1, Martian sand and drift have densities of 1000 kg/m$^3$ to 1300 kg/m$^3$ while crusty to cloddy soil and blocky indurated soil have higher densities of up to 2000 kg/m$^3$. JSC-Mars 1 has a low dry density of around 840 kg/m$^3$ while the other simulant material dry densities span the expected range. It should be noted that recent interpretation of gravity and other geophysical data as well as petrological and geochemical data from Mars and from Martian meteorites (e.g., Belleguic et al., 2005; Pauer and Breuer, 2008; Baratoux et al. 2014) suggest that Martian crustal rock has densities of 2800 to 3200 kg/m$^3$, with extremes ranging up to 3700 kg/m$^3$. Similar values should hold for the basaltic crust at the InSight landing site (Golombek et al., 2020). These values are larger than the bulk density values used by e.g., Herkenhoff et al. (2008) for estimating the density of Martian sand of 1000 to 1500 kg/m$^3$. Thermal measurements using the TEM-A hardware on the mole (Grott et al., 2021) suggest a bulk density for the top 40 cm of regolith under InSight of $1211^{+149}_{-113}$ kg/m$^3$. Using 3200 kg/m$^3$ and 1200 kg/m$^3$ as representative values for the specific gravity of the basalt and the bulk density of the soil, respectively, a porosity of about 63% is calculated. Density and porosity matter for the soil pressure gradient in the Martian regolith and thus the friction on the mole wall and for the resistance to penetration (e.g., Rahim et al., 2004; Poganski et al. 2017, Zhang et al., 2019)

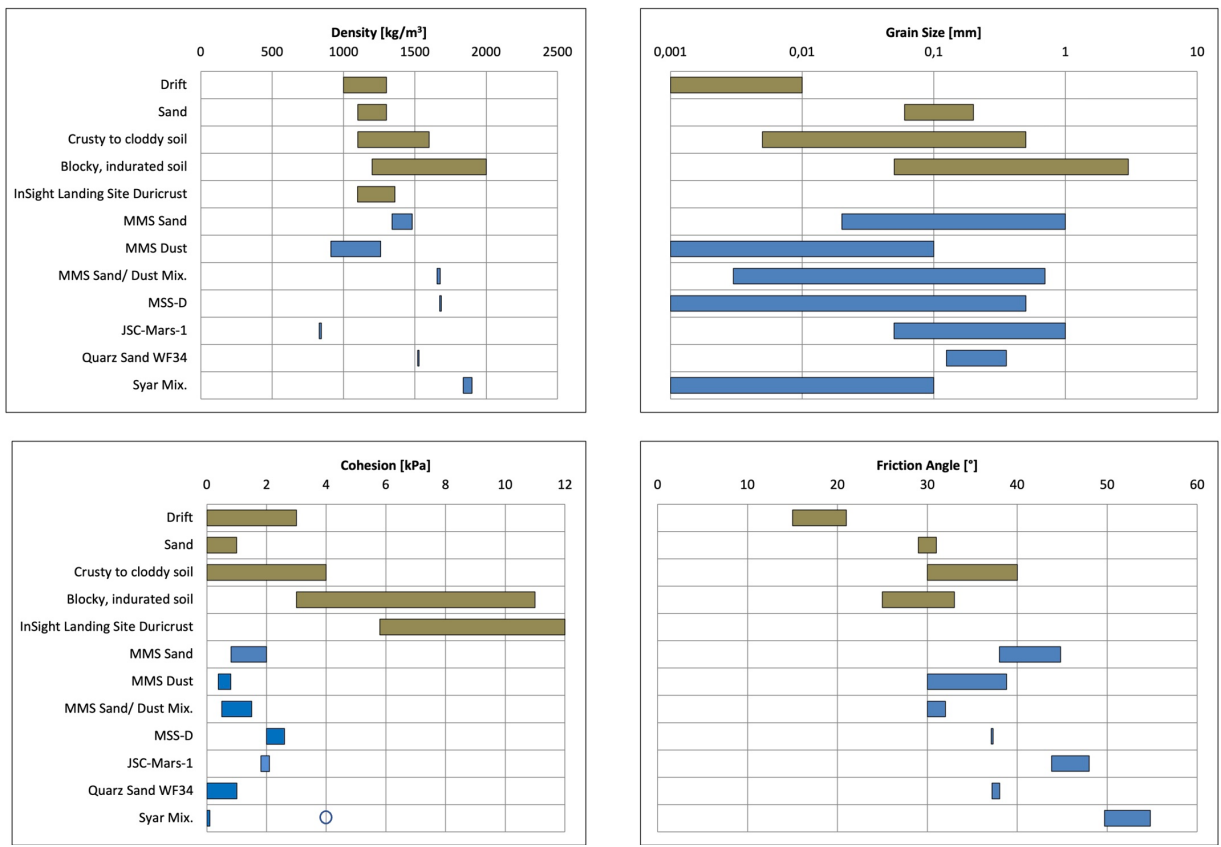

Figure 1 Range of densities, grain sizes, cohesion and friction angles of Martian soil (compiled by Golombek et al., 2008 and Herkenhoff et al., 2008 from various sources) in comparison with those of Mars simulant soils. In additon, the cohesion value derived for the duricrust from slope stability analysis of the walls of the mole pit at the InSight landing site (Marteau et al. 2021; Spohn et al. 2021) is given. The data on Mojave Mars Simulant (MMS) sand are from Peters et al., 2008.

Grain sizes span wide ranges for Syar and MSS-D soils, much wider than are expected for Martian soils, in particular for Mars sand. Qz-sand has a narrower range of grains sizes than e.g., Syar although still somewhat wider than estimated for Martian sand. Martian sand has low cohesion of ≤1 kPa. Drift, crusty to cloddy soil, and blocky indurated soil have higher cohesion up to 11 kPa. Slope stability analysis of the mole pit walls at the InSight landing site gave a cohesion for the duricrust of 5.8 kPa assuming a friction angle of 30° (Marteau et al., 2021; Spohn et al. 2021). From the HP$^3$ mole penetration resistance Spohn et al. (2021) found a cohesion between 4 kPa and 25 kPa for friction angles between 30° and 40°. Qz-sand is basically cohesionless (e.g., Klinkmüller et al. 2016) and thus compares well with Mars sand. A 75/25 Syar sand/dust mixture with 100 $\mu$m Syar dust was used for some Deep Penetration Tests. Triaxial shear tests done at Braun Intertec gave a cohesion of 0.06 kPa, a value that could be confirmed by our independent study using a 80/20 mixture. A direct shear test done at Braun Intertec gave a value of 4 kPa which is marked by an open circle in Figure 1. It should be noted that direct shear tests tend to give higher cohesion values. MSS-D is mildly cohesive with 2 kPa .

Friction angles of 30° can be expected for Mars sand while smaller values have been estimated for Mars drift. Simulants have larger friction angles ranging from 30° to more than 50° for Syar-mix. The compilation above clearly shows that friction angle and size range of particles are the major difference between the soil parameters estimated for Mars and the simulant sands. Mars sand has a limited width of the grain size spectrum while Syar mix and MSS-D have a wide variation. Syar mix was intentionally mixed with 20% dust sized particles, which has not been observed in soils on Mars (e.g., Goetz et al., 2010; Pike et al., 2011). Syar sand is angular as it is produced by fragmentation of basaltic rock. The low thermal inertia observed at the InSight landing site limits the cohesion to values of only a few kPa or less (Piqueux et al., 2021). All Studies report equant to very equant spherical grains and subrounded to rounded to very rounded grains for Mars in contrast to the simulants (e.g., McGlynn et al., 2011; Goetz et al., 2010, Minitti et al., 2013, Ehlmann et al., 2018). Soils with angular shapes may show small cohesion via interlocking grains as do mixtures of sand size particles with finer dust. The leading hypothesis (McGlynn et al. 2011) is that Martian sand is impact generated and is rounded by eolian processes. Moreover, saltation impacts would continually put dust into suspension. From these arguments, Quartz sand "WF 34" and basaltic "MMS <2mm" were chosen as representative test sands.

Penetration tests in some varieties of sands have been performed under ambient conditions at the test site at DLR Bremen in a 5 m high, 80 cm diameter cylindrical container and also at JPL in the 1.7 m high, 60 cm diameter Geothermal Testbed (GTB). The GTB further allowed temperature and atmospheric pressure to be adjusted to Mars conditions. In addition, tests were run in a vacuum chamber at SRC Warsaw (58 cm high, 50 cm diameter) and in a vacuum chamber at ZARM in Bremen (70 cm high, 30cm diameter). The tests in Bremen have been reported in detail in Wippermann et al. (2020) who discussed the results together with the results from the penetration tests at JPL and SRC. In addition to observing the mole penetrating, a calibrated penetrometer was used to measure the penetration resistance in some tests for Qz-sand. The resistance was found to increase with overburden and after compaction due to mole operation but was generally <400 kPa before mole penetration and <1.2 MPa after penetration indicating densification. The mole penetrated at rates significantly larger than the required minimum rate up to 3 m depth, reducing to lower rates at deeper depths in the only chamber (Bremen) that could accommodate such depths. The rapid decrease of the penetration rate in Qz-sand tests at depths >3 m is attributable to the effect of friction on the tether and possibly some effect of the container walls. Nevertheless, the rates were still considered acceptable. In Syar, the mole penetrated significantly more slowly but performed better at depth beyond 3 m with rates in excess of the required 0.25 mm/stroke.

Previous missions to Mars have revealed that the top few cm of the Martian soil could be cemented to form what has been termed a duricrust. Thermal inertia data at the InSIght landing site show no evidence of a duricrust (Figure 2). The low thermal inertia corresponds to a surface dominated by sand size particles with little cohesion. The lack of pronounced seasonal variations in the thermophysical properties argues against a steep inertia contrast produced by cemented material within the top few tens of centimeters (Golombek et al., 2017, 2020).

The mechanical strength of the duricrust, provided by cementation or simply cohesion between regolith particles (rather than, for instance, angular particle shapes) could reduce the lateral soil pressure on the mole in the crust compared to unconsolidated (i.e., cohesionless) material. However, observations of Martian duricrust acquired during previous landed missions suggested that the duricrust would be no more than a few cm thick and would overlie loose sand with properties resembling those of Qz-sand (Herkenhoff et al., (2008) and Golombek et al. (2008). The mole and its support system were thus designed to be able to punch through a weak and shallow crust which, based on the experience from previous missions, was not expected to pose a problem.

As evidenced by the steep walls of the pit in which the mole was found after the support structure had been removed on Sol 227 (Figure 3), the regolith at the InSight landing site, contrary to expectations, did indeed show high cohesion, possibly caused by a cemented duricrust (Golombek et al., 2020). The crust at Homestead hollow is much thicker - at least 7 cm from visual inspection of the pit - and may be even as thick as the final depth to the tip of the HP$^3$ mole of 37 cm (Hudson et al., 2020). Spohn et al. (2021) estimate a thickness of the duricrust of about 20 cm from the length of mole extraction during the back-hammering on Sol 325 (compare section A.4 of the appendix). Estimates of the cohesion of the duricrust derived from arm pushes argue for relative high cohesions (Golombek et al., 2020; Marteau et al., 2021)

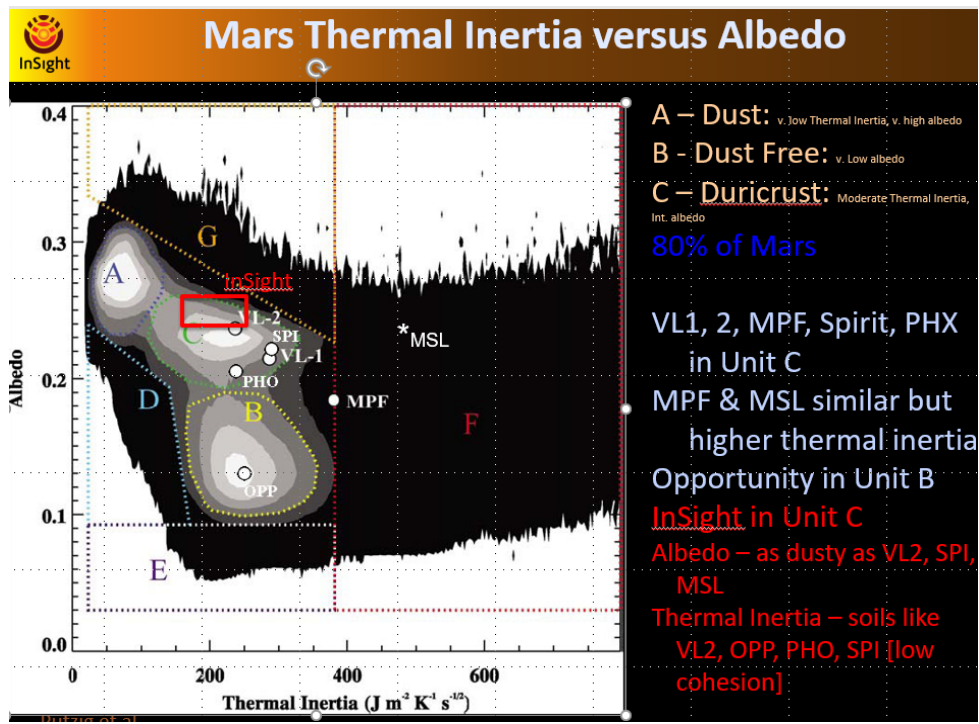

Figure 2. Albedo and thermal inertia for Martian landing sites (see Golombek et al., 2016). The data suggest that the InSight landing site would not have duricrust.

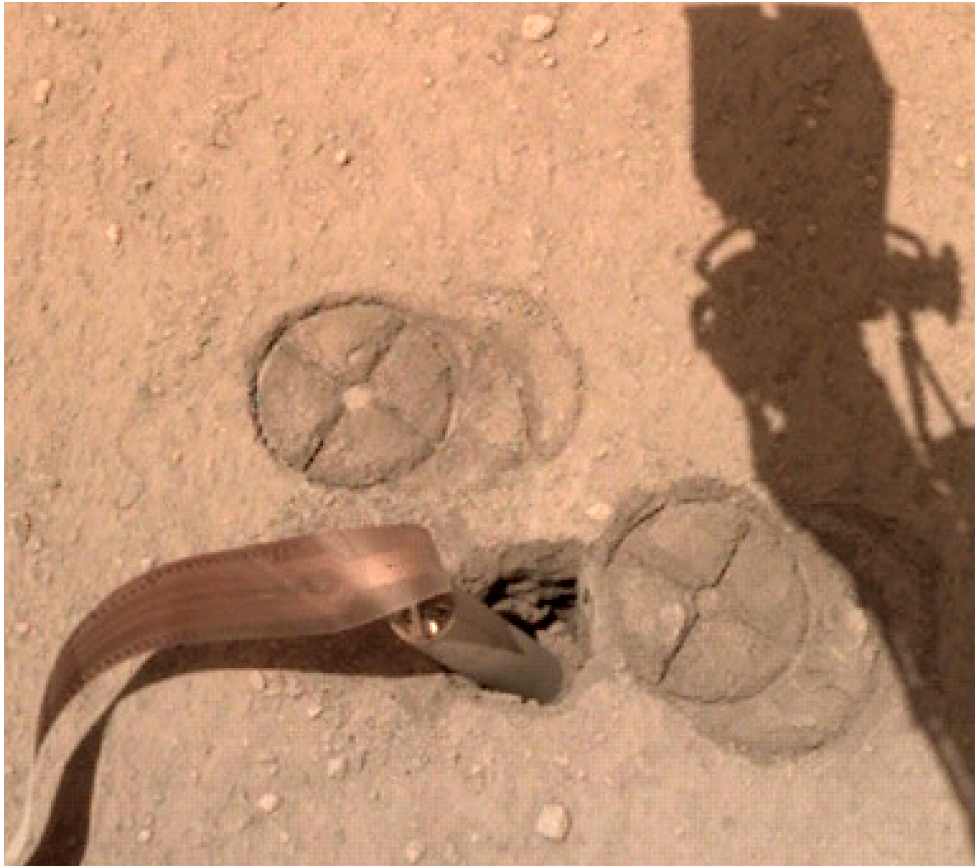

Figure 3. Image of the mole in the pit as seen after removing the support structure on Sol 227. The image was taken on Sol 230. Digital elevation models of the pit show that it is 7 cm deep at its deepest point seen in the image (Golombek et al., 2020, Spohn et al., 2021). The wall is 87° steep.

## 3. The mole penetrator and its support structure, implementation and technical details

The mole is a low velocity penetrator that converts energy stored in a compressed spring to kinetic energy of forward motion. The concept dates to Gromov et al (1997) and was described in e.g., Grygorczuk et al. (2011), Seweryn et al. (2014) and more specifically for the HP$^3$ mole in Krömer et al. (2019) who describe the penetrator in detail. The concept is illustrated in Figure 4. Working as a mechanical diode, momentum is preferentially guided into the direction of planned motion. In the HP$^3$ mole hammer mechanism, a motor provides rotational motion that is converted by a cylindrical cam into translational motion of a piston to compress the drive spring. (Steps 1 to 2 in Figure 4). When the drive spring is released (step 2), its expansion accelerates the hammer that will hit an anvil connected to the casing to transfer the momentum forward (i.e., downward, step 3). At the time of release of the drive spring, a counter-mass, consisting of the motor, gearbox, driveshaft and supporting hardware, is accelerated in the opposite direction to the hammer. This 'supressor-mass' moves upward against the resistance of a second spring (step 4) with a low spring rate, termed the brake spring. Following the compression of this brake spring, it accelerates the suppressor-mass forward (step 5) to provide a second hit to the casing. Figure 5 shows in more detail how the mechanism is arranged in the mole.

To work itself into the soil the mole must deform the soil plastically, and this is done by the high-energy forward stroke. Porosity of the sand is important, as there needs to be enough pore spaces in the soil around the mole, which are collapsible by the hammering strikes or allow the sand to flow. There is another essential component to forward progress: external forces are required to balance the small (relative to the forward drive stroke) backward (recoil) force transferred to the casing along with the compression of the brake spring. These forces are illustrated in Figure 6. The balance between the recoil forces and the forces moving the mole forward depend on the ratios between the hammer and the suppressor-mass (0.24 for the HP$^3$ mole, compare Table 1), the mass of the casing, and the properties of the springs. For the present mole, a force balancing the recoil of 5-6 N is required. Nominally this is 5.4 N on Mars. Under certain unfavorable conditions a force of up to 6.9 N may be required. This is the maximum force to fully compress the brake spring. One such case is that of the mole hitting a very hard surface, which can cause the suppressor mass to directly hit the casing on the upward motion (compare Figure 4, step 4). A further unfavorable case would occur if the mole was attempting to penetrate at a large (order > 60 degrees) inclination.

| Description | Value | unit |
|---|---|---|
| Drive spring potential energy | 0.7 | J |
| Brake spring and wire helix operational forces | 3.7 - 6.9 | N |
| Compression of the drive spring | 15 | mm |
| Maximum travel distance of the suppressor mass | 23 | mm |
| Suppressor mass | 0.46 | kg |
| Mass of the hammer assembly | 0.11 | kg |
| p1: Hammer-to-Supressor mass ratio | 1:4.2 | |
| p2: Hammer-to Outer Casing mass ratio | 1:2.5 | |
| Mass of the casing and other components of the mole | 0.28 | kg |
| Total mass of the mole | 0.85 | kg |

Table 2. Spring parameters and mass breakdown of the HP3 mole.

The design of the mole assumes that the force balancing the recoil is provided by friction between the soil and the casing. During the initial penetration of the mole, when it is only partially in the ground and there is little friction available from the soil, the recoil-resisting force is provided by a set of friction springs in the vertical tube that houses the mole in the support structure (compare Figures 7 and 8 and Reershemius et al. 2019).

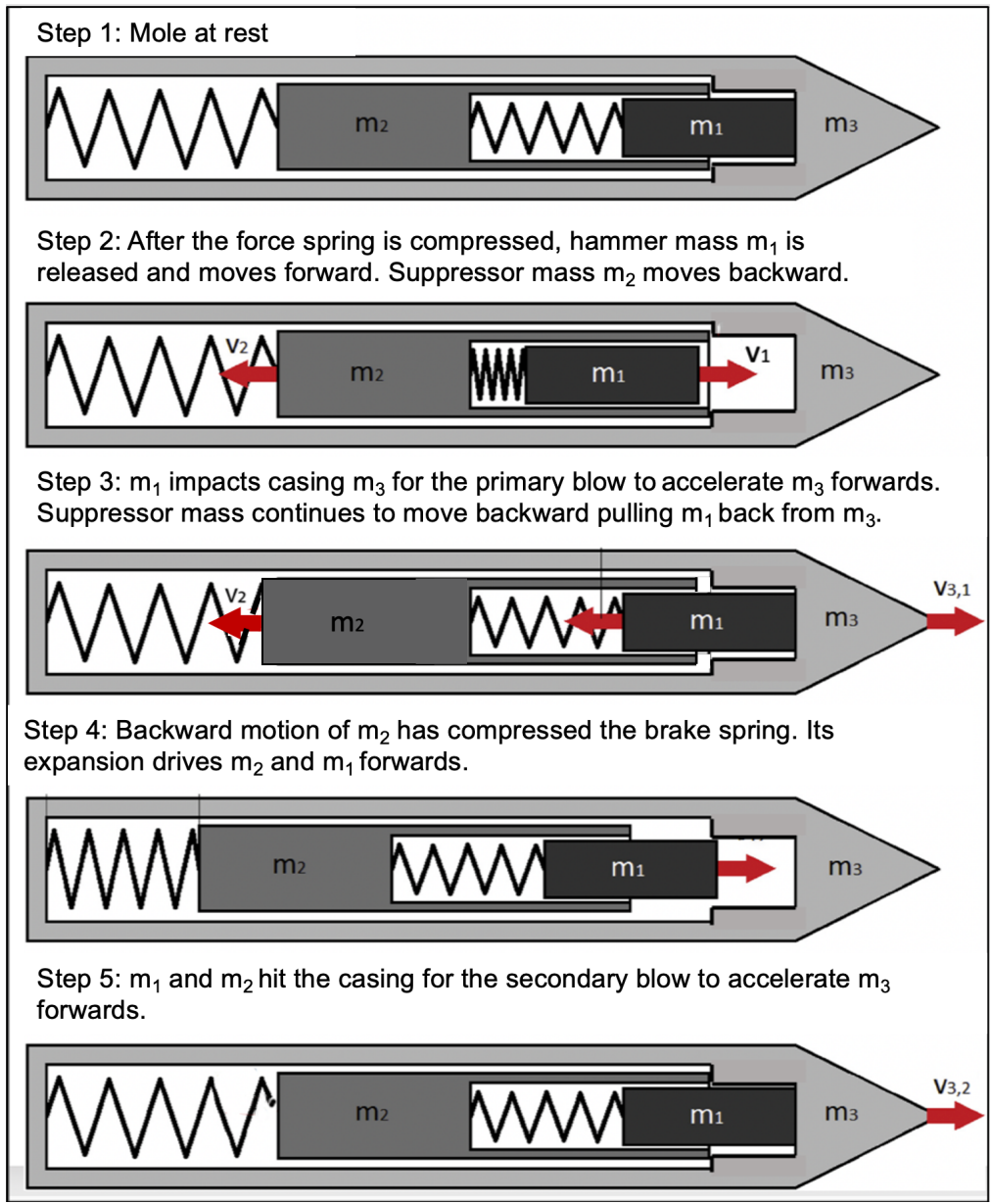

Figure 4. Principles of mole operation illustrated in 5 steps from top to bottom (modified from Krömer et al., 2019) For further explanation see text.

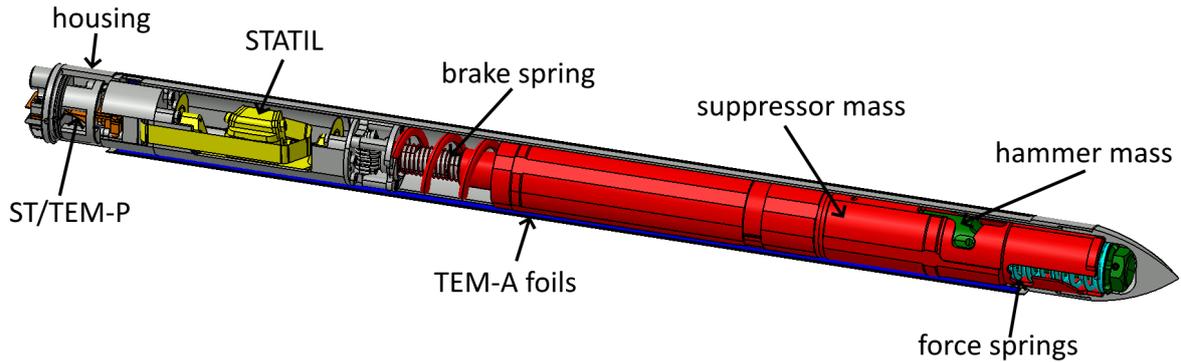

Figure 5. Cross section of the HP³ mole showing the tiltmeter STATIL (yellow), the Science Tether ST/TEM-P attachment (orange), the TEM-A foils (purple), the suppressor mass including the brake spring of the hammering mechanism (red), the hammer mass (green), the force springs (light blue) and the housing (grey).

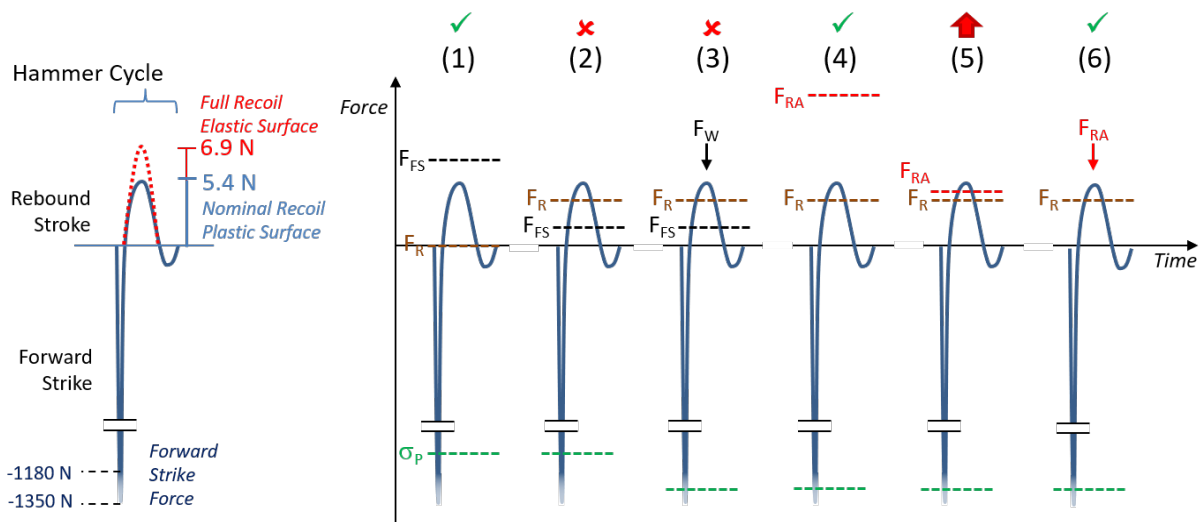

Figure 6 Force-vs-time schematic of a hammer cycle under various circumstances that occurred on Mars. Left panel: Nominal case. Forward strike force is 1180 – 1350 N. The nominal threshold force needed to prevent the outer casing from moving backward is 5.4 N, assuming that the mole moves into a plastically deformable surface that accommodates some forward stroke energy. If the impact surface is hard and elastically returns forward stroke energy into the mole, the rebound recoil can be 6.9 N, representing full compression of the brake spring. Right panel: (1) Mole in SS with full friction spring support ($F_{FS}$) and no regolith friction ($F_R$); (2) Mole in SS with partial friction spring support and some regolith friction, the combined friction resisting the rebound stroke does not exceed recoil force so there is no downward motion; (3) Compaction has increased the plastic deformation threshold ($\sigma_P$) and increased the rebound stroke force (red dotted line), only the weight of the SS ($F_W$) resists mole rebound (but as discussed, the low-mass SS was likely lifted by mole rebounds; (4) SS removed, robotic arm applies pinning force ($F_{RA}$) in excess of rebound stroke, allowing downward movement; (5) Force from Robotic Arm (RA) is insufficiently transferred to the mole, recoil force on rebound stroke exceeds applied forces and a mole reversal event occurs; (6) Back cap push applies pre-load directly along mole axis, resisting rebound and allowing downward movement.

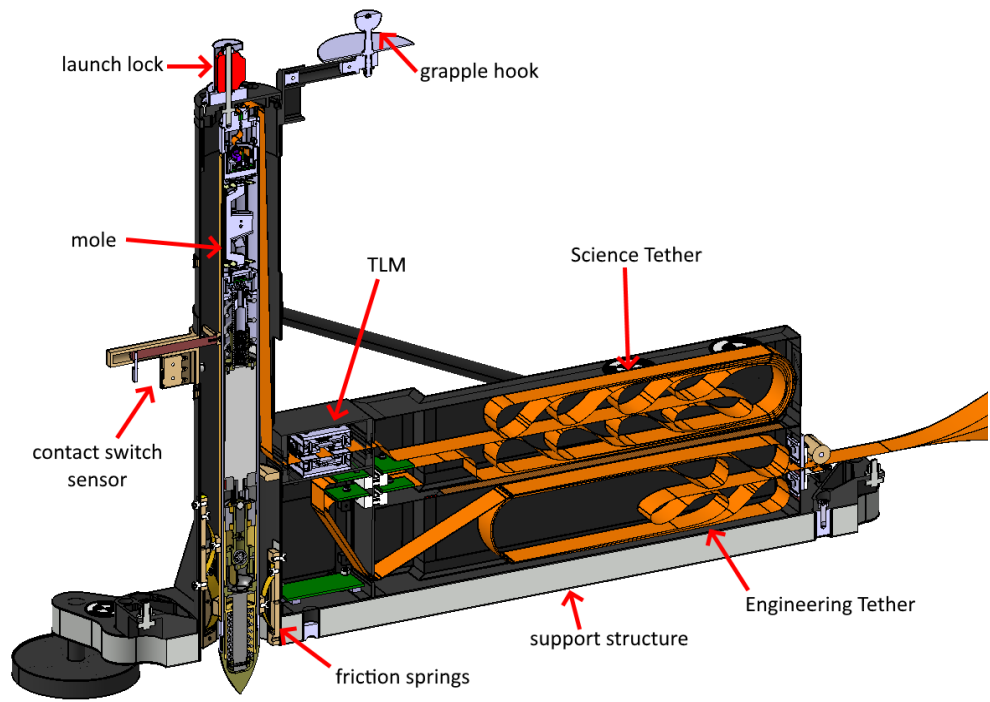

Figure 7 Details of the mole in its support structure made of Carbon Fiber Reinforced Polymer. From left to right, the section shows the right forward foot of the structure. The mole is fixed in the vertical tube by a launch lock at the top and guiding springs near the bottom that provide friction to balance recoil when the mole hammers itself out of the structure. The launch lock was released using a frangibolt before the first hammering. On the upper right of the tube, the grapple hook is shown that was used during HP$^3$ deployment from the lander deck to the surface. At the midpoint of the tube is a spring-loaded contact switch that engages when back of the mole progresses beyond it. The passing of the mole provided the first datum of mole action during the first hammering (compare section A.1 of the appendix). In the box to the right of the tube the tether compartments are shown. The bottom part houses the Engineering Tether (ET) that runs from the support structure to the electronic box on the lander. The upper part is the compartment housing the Science Tether (ST). The latter, embedded with temperature sensors, is connected to the mole and would be pulled by the mole into the ground. The amount of Science Tether drawn from the box is measured by an optical device, the tether-length monitor (TLM). Note the science tether 'service loop', running vertically up the tube, between the TLM and the back of the mole in its launch configuration. Because of this loop, the mole tip must reach a depth of about 54 cm before tether is pulled through the TLM.

The HP$^3$ support system assembly SSA is described in detail in Figures 7 and 8 and in Reershemius et al. (2019). It is a Carbon Fiber Reinforced Polymer (CFRP) structure which hosts all components of the HP³ instrument that need to be placed on the surface of Mars.

Figure 8 shows further details of the guiding friction springs in the vertical tube of the support structure. These springs provide friction when the mole starts to penetrate as it progresses out of the tube into the soil during the initial hammering phase. There are two sets of three springs arranged in two tiers. In each tier, the three springs are located at the same height and spaced at 120° around the tube. The two tiers are in contact with the mole 60 mm and 104 mm above the surface. Each spring is formed from a curved 'leaf spring' of metal fastened at its upper end. The lower end of the spring is curved and rests against a guide block against which it is free to slide. At the apex of each spring is a polymer block, called a 'gliding element' contoured to the shape of the mole's outer casing. See the caption of Figure 8 for more details on the springs' directional behavior.

A contact switch sensor, mounted 248 mm above the regolith surface on the vertical tube, is spring-loaded against the outer casing of the mole. This switch changes state when the mole has moved down in the tube far enough to release the spring mechanism. indicating the moment when the back cap of the mole mole passes this point. This is the only depth indication available during the mole's initial descent. The support structure contained a tether length monitor (TLM) designed to optically read markings on the edges of the science tether, providing measurements of the amount of ST extracted with 2 mm precision. However due to design constraints including SSA stability, available science deck vertical volume, and complexity, the TLM was mounted mid-way down the length of the tube, away from the back cap of the mole at the top. A 'service loop' of the science tether extends from the exit of the TLM up to the back cap of the mole in launch configuration. This loop must be exhausted - at a mole subsurface tip depth of ~540 mm - before new science tether is pulled through the TLM. Only when the tether is pulled through the TLM are new depth data available. This proved to be a major design disadvantage for HP$^3$. Because the mole on Mars penetrated only a distance of about 430 mm measured along the mole axis (37 cm vertical, final inclination 30°), the TLM never engaged, and a reliable estimate of the initial penetration could only be made by recording the time and the number of strokes when the mole passed the contact switch. This occurred 4.6 minutes and 77 strokes after hammering had started. Note that during the mole anomaly efforts, after the support structure had been removed on Sol 227 and if the mole was visible above ground, progress of the mole was measured using stereo imaging (see Spohn et al., 2021).

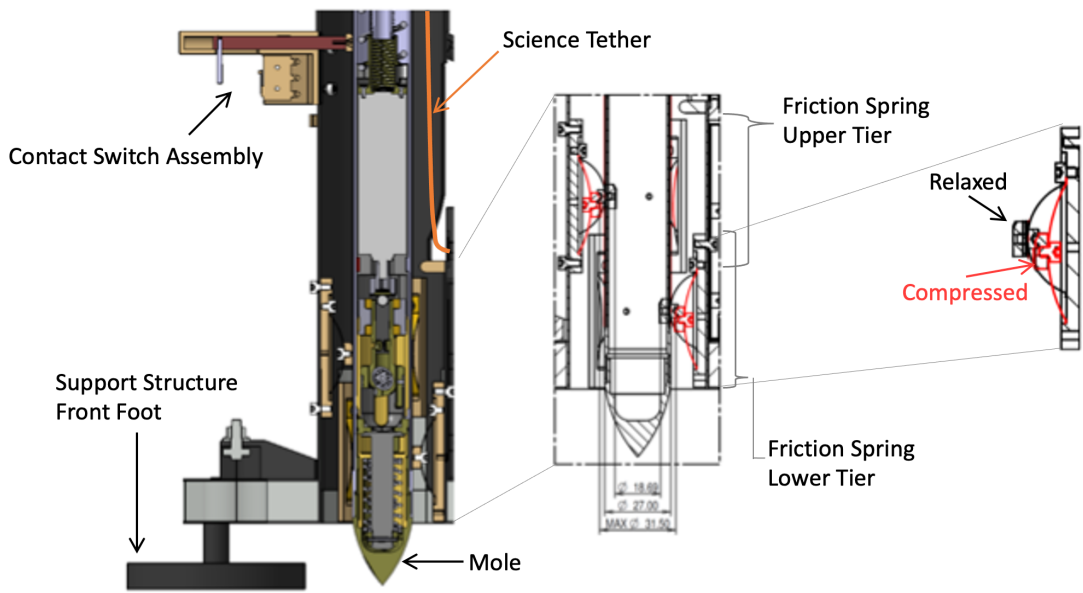

Figure 8. shows details of the arrangement of friction springs near the bottom of the tube (left and middle panel). The contact switch assembly is also visible at the upper left of the figure. In addition to guiding the mole, the springs provide friction to balance the recoil force generated in the hammer mechanism and transferred to the mole casing. The action of the springs depends on the direction of motion of the mole (Right panel). Sliding 'up' is restricted by friction on the gliding element pulling the spring into greater compression, which in turn forces the gliding element further against the mole. The resulting horizontal force is then ~ 30 ± 2 N. Sliding 'down' is accommodated by spring compliance and freedom-of-motion of the spring's lower end and results in ~ 7 ± 0.5 N (measured horizontally).

The mass of the support structure without the mole is 2 kg. On Mars the weight of the structure is thus only 7.4 N, only a little more than the force provided by the friction springs. This is the time when the mole started to penetrate with an increasing inclination-from-vertical of 11° (section A.1 of the appendix). Considering the data recorded by the STATIL sensors during this time, and the geometry of the support structure, it is concluded (Spohn et al., 2021) that the mole hammering action lifted the support structure away from the regolith surface during at least the first 77 hammer strokes, due to a ratcheting motion against the friction springs.

In principle (and as seen during some inclined penetration tests at DLR Bremen) because of its low center of gravity, the mole tends to align itself with the vertical if it able to penetrate several mole lengths. But this did not occur during the small distance of penetration (about 1 mole length) that the mole finally achieved. Images taken on Sols 118 and 158 (during the so-called 'diagnostic hammerings' described in section A.2 of the appendix) showed that the mole jostled the support structure. The resultant movement was also revealed by footprints of the support structure feet in the sand (Figure 3). The unintended horizontal and especially vertical motion of the support structure during the initial penetration and anomaly resolution steps negatively impacted the mole's ability to penetrate.

**Lessons learned:**

**1. The transition of the mole from its above-ground state in the support structure to a fully buried state is critical! To evaluate and react to initial penetration behavior, there needs to be a reliable and high-resolution (both in space and time) measurement of the mole's progress during the initial phase, at least as accurate as that planned for the later penetration phase. Other helpful diagnostic tools could include a sensor to measure movement of the support structure.**

**2. Movement of the support structure is detrimental to initial penetration performance and should be minimized by anchoring features, increased weight, and/or applied loads (e.g., from an arm). Securing the structure in place would have better guided the mole during initial penetration and may have avoided the mole penetrating at an increasing tilt angle.**

## 4. Force Balance Considerations. Penetration Resistance and Hull Friction

4.1 Penetration Resistance and Rate

The HP$^3$ mole flight model has a constant drive spring potential energy of 0.7 J. When released, the energy results in an impact force between 1180 and 1350 N for a period of less than 0.1 msec. These values were measured in a test stand in which the mole was hammering against surfaces varying from rigid to supported by a spring with a 1852 N/m spring constant. They are equivalent to vertical stresses of 2.1 to 2.4 MPa, greater by a factor of about two compared to the cone-penetration resistances measured by Wippermann et al (2020) of the sands in the 5 m tall test chamber. Rewriting Equation (2) using the known energy of the HP$^3$ mole drive spring and its diameter of 27 mm, we get

33) $$\sigma_P \cdot R \approx 1.22 \cdot 10^3 \cdot \varepsilon \text{ mm/stroke}$$

Thus, the HP$^3$ mole could have been expected to penetrate a material with MPa resistance at a rate of $\varepsilon$ mm/stroke. The efficiency, $\varepsilon$, of the mole is more difficult to estimate and depends on details of the design of the hammer mechanism. In particular it depends on the mass ratios p1 and p2 (Table 2), and likely on the soil rheology and other parameters quantifying dissipation due to friction in the mole. The ratio between the spring energy and the mole kinetic energy of forward motion calculated from the chosen spring parameters and mass ratios is 0.53, for a coefficient of restitution of 0.9. From the results of the Deep Penetration Tests at DLR reported by Wippermann et al. (2020) an efficiency of $\varepsilon = 0.47 \pm 0.05$ could be estimated (Spohn et al., 2021). This value is consistent with estimates from discrete element modeling of about 0.4 - 0.5 (Zhang et al. 2019) while field tests give 0.3 - 0.9 as these authors report. It is smaller than an estimate from the analytical cavity expansion model of Salgado et al. (1997) for which Rahim et al. (2004) give 0.75. Zhang et al. (2019), citing evidence from Discrete Element Modeling as well as from field measurements, report that the penetration resistance increases with the square of the relative density and linearly with overburden pressure. Rahim et al. (2004) find the penetration resistance to depend on initial porosity, friction angle of the granular material and dilatancy, while cohesion was found to be of smaller importance. For small friction angles of 20° or less, the dependence on porosity is small, with resistance increasing by a few percent when porosity is decreased from e.g., 0.6 to 0.3. Resistance will increase by a factor of 5.5 though in that same porosity range for a friction angle of 40°. Poganski et al. (2017) using an analysis by Terzaghi and Peck (1947) find the penetration resistance to linearly increase with depth but to depend exponentially on $\tan\phi$. Comparing Qz sand and Syar mix, the penetration resistance of the latter at the same pressure should be a factor of 4-6 larger, which is consistent with the results of the tests of Wippermann et al (2020). The data collected in Figure 1 show that Martian soils have friction angles between 30 and 40° with drift sands having friction angles as low as 20°.

Mohr-Coulomb-Rankine theory as applied to soils (e.g., Verruijt, 2018) relates the principal stresses $\sigma_1$ and $\sigma_3$ ($\sigma_1 > \sigma_3$) at failure to the angle of internal friction $\phi$ and cohesion $c$ via

4) $$\sigma_1 = \frac{\sigma_3}{K} + \frac{2c}{\sqrt{K}}$$

where

5) $$K = \frac{1-\sin\phi}{1+\sin\phi}$$

The lateral passive earth pressure (resistance) is obtained by setting $\sigma_3 = \rho g z$

6) $$\sigma_{P0} = \frac{\rho g z}{K} + \frac{2c}{\sqrt{K}}$$

$\sigma_{P0}$ is the stress against which the cavity containing the mole needs to be radially expanded when the mole moves forward - ignoring the expansion at the tip.

Equation (6) recovers the linear dependence on overburden reported by Zhang et al (2019) but not the dependence on relative density, change of porosity upon penetration, dilatancy, and penetrator geometry (Rahim et al. 2004). More detailed analytical and numerical treatments find that the resistance to penetration is easily an order of magnitude larger than $\sigma_{P0}$. This is consistent with measurements of the penetration resistance in the Deep Penetration Testbed in Bremen (Wippermann et al., 2020) for Qz-sand which is about an order of magnitude larger than $\sigma_{P0}$.

The penetration rates observed on Mars were significantly smaller than in the tests on Earth. While in the Deep Penetration Tests, penetration rates were found to be several mm/stroke at very shallow depths even in compacted and cohesive sands, the initial penetration rate on Mars was estimated as 0.5-1.2 mm/stroke. At tip depths >31 cm, the rate was observed to be no greater than 0.11 mm/stroke (Spohn et al., 2021). In the terrestrial testbed at depths of 30-40 cm in Qz-sand, the rates were about 2.4 mm/stroke on average. Note that because of the lower gravity on Mars a depth of 30 cm on Earth has the same overburden pressure as a depth of about 80 cm on Mars. Using Equation (3), the penetration resistance on Mars was 0.5 to 1 MPa at shallow depths in the shallow duricrust, but >5 MPa at tip depths of 31 cm to 37 cm, the latter being the final tip depth of the mole. Penetration resistances in the Deep Penetration Test bed were 0.2 to 0.5 MPa at the equivalent depths.

The reason for the large penetration resistance on Mars is not known. Two prominent possibilities include (1) the strength of the cemented duricrust, which may explain part of the resistance, as well as (2) possible compaction of the sand at a depth of 31 cm, the depth where the mole tip executed roughly 8700 hammer strokes during Sols 92 and 94 (see section A.1 of the appendix). The large number of hammer strokes may have caused vibrations that may have added to a compaction of the soil. There is evidence of the mole having precessed about a point approximately midway along its hull thereby widening the cavity (Figure 3) which would have reduced friction on its hull (Spohn et al., 2021). Another, potentially concurrent, possibility is a layer of gravel or pebbles embedded in the sand. Nagy et al. (2020) have done thermal measurements including temperature logging in boreholes in Mars-analog sites in the Atacama Desert (Ojos del Salado, Puna de Atacama). B. Nagy reported in a personal communication (2021) about the failure of penetrating with a hammer-driven rod deeper than 40-50 cm. They found the regolith to consist of 5-10 cm size stones imbedded in fine grains. Penetration was possible when the matrix was loosened using e.g., a rotating cross-bit drill.

The cohesion of the duricrust at the mole pit was estimated using slope stability analysis after the pit wall had partially failed due to pushing the blade of the scoop onto the surface and resulted in a value of 5.8 kPa at an assumed internal friction angle of 30° (Golombek et al., 2020; Marteau et al., 2021; Spohn et al. 2021). This value is comparable to but exceeds the strongest soils measured on Mars (the strong, blocky, indurated soil at Viking Lander 2) (Moore et al., 1987).

Spohn et al. (2021) have argued for a layer of sand between the bottom of the duricrust (at 7 cm or greater depth) and the surface of the resistive layer at 31 cm depth. The existence of such a third layer was inferred from features in the recording of the STATIL tilt meter sensors and of the InSight Seismometer but remains speculative. The penetration rate in that middle layer would have been about equal to the penetration rate at shallow depths observed in the terrestrial test beds in Qz-sand.

An important reason for commanding the large numbers of hammer strokes on Sols 92 and 94 was to penetrate to the target depth as quickly as possible. The lander shadow perturbs the thermal environment, and the perturbation was expected to expand with time to depth at a level that would start to disturb the thermal measurements after about 200 sols (Grott, 2009). The HP$^3$ team was aiming to outpace the perturbation by getting the mole to depth as quickly as possible.

**Lessons learned:**

**3. The mole should have been made more energetic such that penetration at the required rate of 0.25 mm/stroke or more would have been possible even in MPa resistant soils. An electromagnetic mole mechanism such as in the MUPUS or in the EMOLE penetrator (Grygorczuk et al., 2016; Spohn et al., 2014) could have allowed the hammer energy to be adjusted to changes in penetration resistance.**

**4. Initial penetration should be approached cautiously, even if time constraints motivate fast action, so that unexpected hammering-in-place does not create adverse effects such as densification.**

4.2 Friction on the mole hull and failure to penetrate

As has been discussed above, friction on the mole hull is required to balance the recoil of the hammer mechanism. We illustrate the problem of the lack of sufficient frictional resistance in c-$\phi$ soils by considering the horizontal stress in the soil. An empirical estimate of the pressure that would act on the wall of a cavity after the cavity has been opened is

7) $$\sigma_{3a} = \rho g z \cdot K - 2c \cdot \sqrt{K}$$

The active soil pressure is the pressure that is extended to the wall of the cavity as the material surrounding the wall outside the cavity starts to collapse. It can be easily recognized how cohesion decreases the active soil pressure to zero at a certain depth. Applied to the mole, this simply means that depending on the cohesion of the soil, hull friction may be insufficient or even unavailable to balance the recoil up to a certain depth $z_0$. Moreover, mole vibrations may even open a cavity around the mole up to that depth and may leave part of the mole without contact to the soil at all (Figure 3). Using Equation (7), the "zero-active-pressure" depth $z_0$ can be calculated up to which cohesion compensates the overburden pressure:

8) $$z_0 = \frac{2c}{\rho g \cdot \sqrt{K}}$$

Figure 9 shows lines of predicted constant $z_0$ as a function of cohesion and of $\rho g \cdot \sqrt{K}$. The isolines mark all soil parameter combinations that result in the same thickness $z_0$ (in meters) of the surface layer without active soil pressure. Data points in the chart refer to Mars analogue sands used during the mole Deep Penetration Test activity described in section 2.3 of this paper and in Wippermann et al. (2020). The data point marked "InSight Landing Site" uses the cohesion value derived for the duricrust in Spohn et al. (2021) updated from earlier data in Golombek et al. (2020) and Marteau et al. (2021).

Even for cohesion values of only a few kPa, $z_0$ can exceed a mole length. Only for cohesionless sand would friction be available at very shallow depths. Analogue materials used in terrestrial test-beds thus should be of higher cohesion than expected on Mars to compensate for the higher Earth gravity. For MMS-Sand on Earth, a similarly thick low-active-soil-pressure layer is expected as for mildly cohesive sand on Mars (refer to the $z_0$ = 0.5 m isoline in Figure 9). Accordingly, for the insertion phase and its low soil pressure regime, MMS-Sand was an acceptable analogue to Mars sand. The area to the left of the $z_0$ = 0.5 m isoline can be regarded as representing the soil parameter ranges for which a successful mole insertion could be expected based on the test results reported in Wippermann et al. (2020) and the assumption that Mars Sand with properties as in Figure 1 (drift, sand, and crusty-to-cloddy soil) would be representative of the Martian soil. The conditions found at the InSight landing site and reported in Marteau et al. (2021) and Spohn et al. (2021), which fall on the $z_0$ = 4 m isoline, are significantly off the verified domain. Further consequences from these considerations on the mole's verification and validation campaign are discussed in section 5 below.

It should be noted, though, that Figure 9 may overestimate the thickness of the low friction zone. Local failure of the soil that may cause an earlier onset of active pressure at smaller depths than predicted by Equation (8). The calculation also assumes the soil to be homogenous and without stratification.

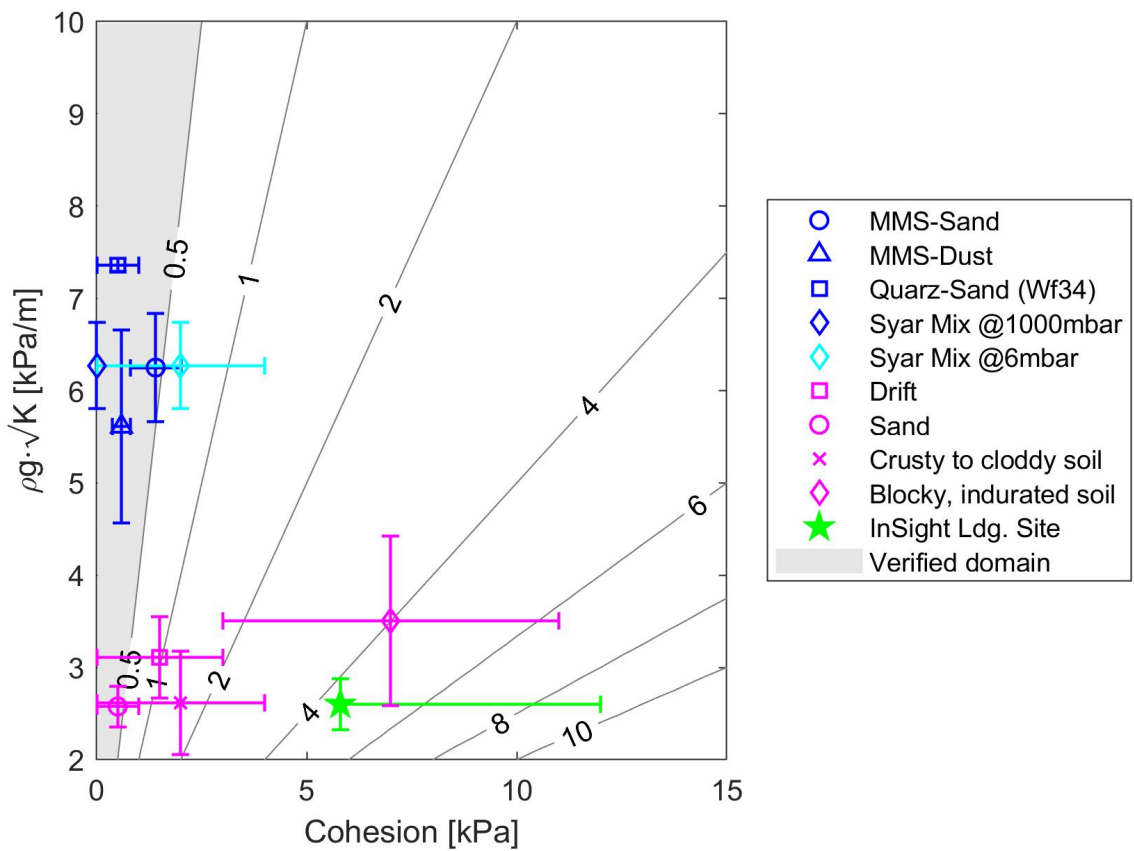

Figure 9. Lines of constant zero active pressure depth in meters as predicted by Equation (8) as a function of cohesion c and $\rho g \cdot \sqrt{K}$. The plot includes data from the test sands used in the Deep Penetration Tests (MMS, qz-sand Wf 34, and Syar sand, blue data points) and from various Mars mission landing sites (magenta data points as reported in Golombek et al., 2008). The entries are the mean values of the ranges given in Figure 1. The cohesion for the InSight landing site is the value estimated using slope stability analysis in Marteau et al. (2021) and updated to include the results of numerical calculations in Spohn et.al (2021). Also given is an estimate of the Earth atmosphere pressure effect for Syar sand. For further explanation of the pressure effect, see section 4.3.

## 4.3 Atmosphere pressure effect

An unexpected effect that can affect mole penetration was discovered during tests of the mole in sands at low atmosphere pressure. Penetration tests at DLR Bremen and at JPL have shown that the Proto Flight-Equivalent Mole model 1, PFE-1, (a mole model that was technically equivalent to the Flight Model) failed to penetrate at shallow depth in Syar sand and MSS-D sand at a pressure of 6 mbar (as on Mars) in the Geothermal Testbed. At these conditions the mole progressed for a few hammer strokes before it began to jump up and down with no net forward movement. In some cases, the mole even moved backwards. It made no difference whether the mole was inserted to about half of its length by hand, or was released from a friction-providing deployment device after having penetrated to half of its length. The unexpected behavior was noticed when the ambient pressure was increased while the mole was still hammering in this 'bouncing' state. As the pressure was increased, the mole would again begin to make downward progress when the ambient pressure reached between 60 and 100 mbar. The effect was observed to be reversible, with penetration occurring whenever the chamber pressure was raised to or above this range but returning to the bouncing state when the pressure was dropped closer to Mars surface pressures.

Systematic variation of the pressure in the Bremen vacuum chamber showed that the PFE-1 mole would penetrate in MSS-D at pressures above 100 mbar but not at smaller pressures. Further investigation showed that in a cohesionless sand, such as Qz-sand in the Bremen facility, the surrounding pressure had no effect and the mole penetrated equally well at 6 mbar, 100 mbar, and at ambient pressure. It should be noted that the mole body was vented through a sintered metal frit and thus would achieve the same internal pressure as its ambient environment (without allowing dust to infiltrate and jam the mechanism. Furthermore, the internal design of the hammering mechanism incorporated airflow paths to prevent the negative 'air springs' effects. Force measurements using the test stands in vacuum chambers showed that the mole's forward and rebound forces were unaffected by ambient pressure levels.

The observations of this pressure effect can be understood if the external forces on the mole are considered (compare Figure 10). The Earth atmosphere pressure of 1 bar provides a force of about 57.3 N on the upper circular surface of the mole's back cap. When the recoil occurs during mole hammering, upward movement of the mole will open a low-pressure cavity near the tip of the mole. In a circumstance where this cavity is isolated from the ambient pressure, i.e., there are no large pathways along the mole casing connecting the tip cavity to the atmosphere, gas can only flow into the cavity through the porous soil. If such a pressure imbalance can exist, the mole's recoil force must work against the atmosphere pressure force. The Earth atmosphere pressure force easily surpasses the maximum recoil force of 6.9 N for the HP$^3$ mole even in cohesive soils with zero active soil pressure, allowing progress to be made in such soils. At an ambient pressure of about 100 mbar, the recoil force is just balanced. For ambient pressures below about 100 mbar, the pressure force imbalance drops below the recoil force threshold and mole starts to bounce in sufficiently cohesive soil. The 1-bar Syar datum in Figure 9 indicates how atmosphere pressure on Earth can compensate the effect of cohesion on the force balance.

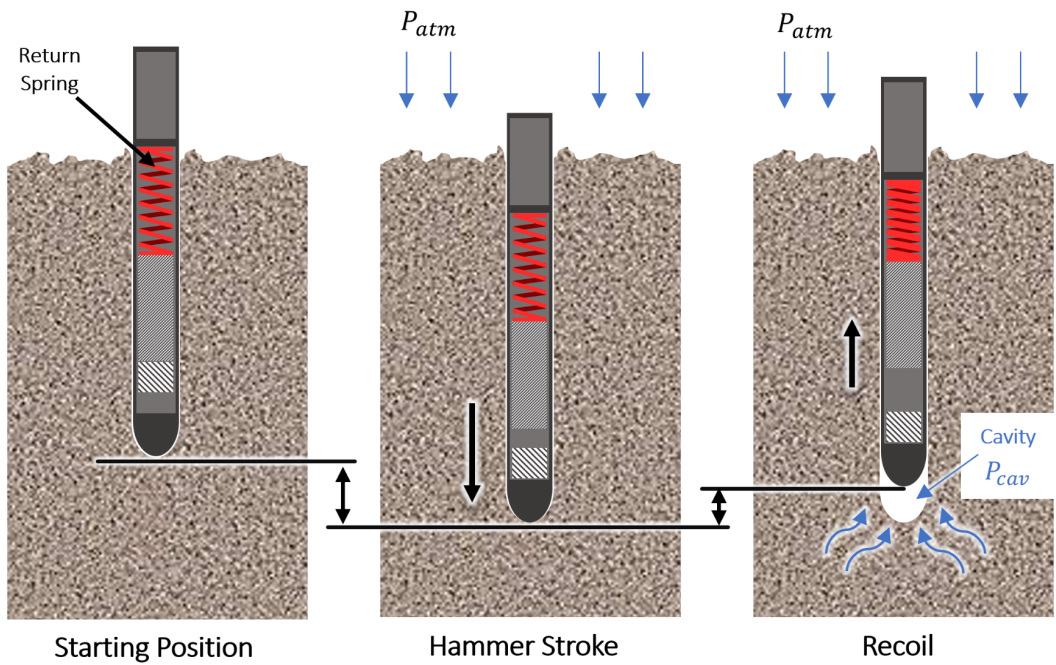

Figure 10. Atmosphere pressure effect on the mole force balance. When the recoil from the hammer mechanism and from the elastic response of the soil force the mole to move upward, a low pressure cavity opens around the tip. The pressure difference between the atmosphere and the cavity will cause a force in the downward direction which can compensate the recoil depending on the pressure difference, provided that the gas flow in the soil (i.e., its gas permittivity) is slow compared to the cavity expansion.

For Qz-sand, we argue as follows: Assuming $\frac{d\sigma_h}{dz} \approx 0.5\rho g$ as suggested in simple empirical estimates for cohesionless sand, the horizontal stress on the hull averaged over the length of the mole is about 15 mbar. Integrating over the surface of a cylinder with radius of 2.7 cm and a length of 40 cm, and using a friction coefficient of 0.3 as appropriate for steel on rock, we find a friction force of about 15 N, sufficient to balance the recoil force. On Mars, using the same simple estimate we arrive at a friction force of about 6 N, or 9 N if the atmosphere pressure of 6 mbar is accounted for. This explains why the mole worked for cohesionless soil at low pressure in the test beds and suggested that it should have worked on Mars for a mole fully buried in cohesionless Martian soil. Please note that a friction coefficient of 0.3 is a conservative estimate.

Even though the force balance estimates suggested that the mole should penetrate on Mars after leaving the support structure and entering into the soil, the experience from Mars show that it did not work this way. The reason was that the Martian soil was not well represented by the Qz-sand simulant that was used for the tests.

**Lesson learned:**

**5. Balancing the recoil from the hammer mechanism is critical during the transition from the support structure to fully buried in the soil. In addition to the hammering mechanism being optimized to minimize the recoil, a mechanism to support the mole penetrating for 1-2 mole lengths should be considered, if only to provide margin. In cohesive sand, the depth must be such that the wall pressure due to the overburden is large enough to balance the recoil.**

## 5. Mole Verification and Validation

The requirement associated to the mole's penetration performance L5-TM-4 (Table 1) calls for the mole to be capable of reaching 3 m depth under Earth ambient conditions into representative Mars simulant ("MMS<2mm" and "Wf-34") within 24 hours of cumulative hammering. An earlier version of the requirements had specified Syar as a third soil simulant. It was later removed as Syar was classified as "not representative" of Martian soil mostly because its angular particles yielded higher friction angles than rounded sand observed in Mars soils (e.g., McGlynn et al., 2011; Goetz et al., 2010, Minitti et al., 2013, Ehlmann et al., 2018). Generally, lower tier requirements are cascaded down from overarching mission and system requirements. The selection of "representative soils" followed assessments and the selection of landing site candidates (Golombek et al., 2017). Their properties flowed down into the respective instrument and mechanisms requirements.

The objective of verification is then "to demonstrate through dedicated processes [test, analysis, review of design, inspection] that the deliverable product meets the specified requirements" ([ECSS-E-ST-10-02C, similar also in the NASE Systems Engineering Handbook). A comprehensive description of the mole deep penetration test campaigns is given by Wippermann et. al (2020). Those tests demonstrated the mole's penetration performance into the specified soils as required by L5-TM-4. From a formal system engineering point of view, it was correctly stated that the mole system was successfully verified. The complimentary objective of validation is to give proof that the system can accomplish its purpose, here: penetration into the Martian surface to a depth greater than 3 m. Validation thus ensures that the requirements and the system build upon them are "the right thing" for the task. In case of the mole, in retrospect, it must be concluded that it was insufficiently validated.

However, an inherent characteristic in any case of exploration technology is that it not just has to deliver a function and performance in a planetary environment but has to interact with that environment. The environment, in this case the soil and its describing parameters, are an active constituent in the functional relation which interacts and is being altered itself by the function's execution, e.g. compressing and displacing soil. It must be kept in mind that ultimately the system (the instrument) and its interactions with the planetary environment, cannot be fully validated upfront (Lorenz, 2018). This renders validation in planetary exploration particularly challenging.

Another perspective is that of agnosticism. While undertaking ambitious endeavors, a certain faith or optimism is essential. A disciplined view of development should consider that a given design or test condition does not have a 100% chance of success, simply because no system has a 100% chance of success, nor for that matter of 0% (sometimes referred to as 'Cromwell's Rule' - see, e.g. Lorenz, 2019). This perspective forms the basis for Laplace's Rule of Succession, namely that the best expectation value of the reliability of a system, given k successes out of n trials, is not (k/n), but (k+1)/(n+2). In other words, the information gained from tests (k,n) is diluted in a Bayesian sense by the prior information that the next test could go either way. Thus, a single successful test does not indicate 100% confidence, but rather only a better-than-even 66%. The chance for the unmodified mole to achieve its goal of penetration to its target depth in a similar environment on Mars at a next opportunity would thus formally be < 1/3.

**Lesson learned:**

**6. The HP$^3$ mole verification tests were successful, but the requirements were too narrowly focused such that the validation failed in retrospect. Had timing and resources permitted, a wider range of soil parameters should have been considered, with regards to cohesion of the Martian soil, thickness of the duricrust, and penetration resistance of consolidated soil.**

# 6. Discussion, Lessons Learned and Suggestions for Improvements of Mole and Support Structure Design

In the foregoing we have discussed the observations made during almost 2 years on Mars attempting to get the HP$^3$ mole to penetrate to its target depth and have derived six lessons learned from these observations. The problems for the mole to penetrate to its target depth have been analyzed to be twofold:

1. A lack of friction from a layer of cohesive soil that may even have been partially cemented. When the mole initially penetrated into the Martian soil down to 31 cm tip depth, sufficient friction to balance the recoil transferred from the hammer mechanism to the mole body was provided by the friction springs in the support structure, at least during the initial ~half-mole-length of penetration (i.e., to the contact switch). After the mole had left the support structure friction from the Martian soil did not suffice to balance the recoil. This caused the mole to bounce in place while precessing and enlarging a cavity or pit, resulting in about 1/5 of its length having loose or no contact to soil! The absence of further penetration together with the evidence for the mole continuing to hammer (for a total of 8601 strokes) and the evidence for the support structure having been lifted and jostled adds credibility to this conjecture. Contrary to assumptions that entered into the requirements for the design of the mole, a layer of at least 7 cm thickness - but more likely 19 cm (Hudson et al., 2020; Spohn et al., 2021) - of duricrust was found with substantial cohesion. The cohesion was such that the active soil pressure was too small to provide friction to the hull for at least 7 cm (open pit depth) but possibly for up to the entire mole length.
2. In addition to the lack of friction, the mole had encountered a layer of unexpectedly high penetration resistance at a depth of 31 cm. The existence of this layer and its resistance was derived from the penetration rate of the mole when it successfully penetrated some centimeters deeper when the recoil was balanced by the scoop of the robotic arm (Section A.3 - A.5 of the appendix). The mole may have encountered a lens of ejecta from a nearby crater (Golombek et al. 2020), or it may have self-densified the soil during the over 8000 hammer strokes with the tip at the same depth during sols 92 and 94. It is possible that the mole may have penetrated further even with low friction had the soil been less resistant.

The four measures to address these problems for future cavity-expansion penetrators are:

1. Further reduce the recoil of the hammer mechanism to the mole casing.
2. Increase the amount and vertical extent of known friction. Possibly increase the effective friction coefficient of the mole hull and provide a structure to support the mole for more than roughly 4/5 of its length of penetration, perhaps even including assistance to the mole below the regolith surface.
3. Increase the energy available from the mole hammer mechanism and make the energy levels controllable
4. Increase the mass of the Support Structure such that the mole will not be able to lift and jostle the device. This will provide more stable initial penetration and work against mole inclination.

The recoil of the hammer mechanism transferred to the mole casing is largely a function of the mass of the mole and the ratios between the hammer and the suppressor masses $p1$ and the hammer and the casing masses $p2$ (compare Table 2). Seweryn et al. (2014) show how the masses can be optimized for a given total mole mass to provide the maximum forward energy of the mole while minimizing the recoil. It is even possible to reduce the recoil such that it will be insignificant if the weight of the suppressor mass can be made large enough. An increase in the suppressor mass adds to the mole system mass, however. When the hammer mechanism was redesigned in 2013/2014, the InSight payload mass and volume allocations for HP$^3$ required that the mole not exceed 1 kg including margin (the final flight mole had a mass of 850 g) and a length of 40 cm. For that mass and length, the smallest maximum recoil force while maximizing the mole kinetic energy was 7 N. A mole mass of 2 kg would have offered a chance to balance the recoil force by the weight of the suppressor mass almost entirely. Extending the length of the back spring by about 60 mm could have had a similar effect. The 2 kg mole with the same drive spring as HP$^3$ would have been 25% more efficient in turning spring energy into kinetic energy of mole forward motion.

Figure 11 shows how the kinetic energy of the mole would vary with the ratio between the hammer mass and the suppressor mass for a fixed value of the mass of the casing of 0.28 kg and two values of mole mass and spring energy, respectively. The energy of the mole can be doubled by doubling the spring energy at the same total mass of 2 kg, thus keeping the efficiency at 0.7. The latter mole would have a recoil though of 2-3 N. Reducing the recoil to almost zero would require a value of *p1* of 0.08, reducing the kinetic energy to 0.91 J, or a further increase of the mole mass.

It should be acknowledged, though, that an increase in mole mass would imply an increase in mole dimensions, either in diameter or in length, or both. Increasing the mole diameter is unfavorable for the penetration performance as more soil must be displaced. Increasing the length would be favorable at shallow depths, since it would increase available surface area for mole-regolith friction.   At greater depths, however, the increased friction due to greater length would tend to reduce the penetration rate.  Other constraints can limit the maximum length of the mole; for instance in the case of InSight, the mole (and support structure) could not be increased in length (height) because no more vertical space was available under the lander's backshell.

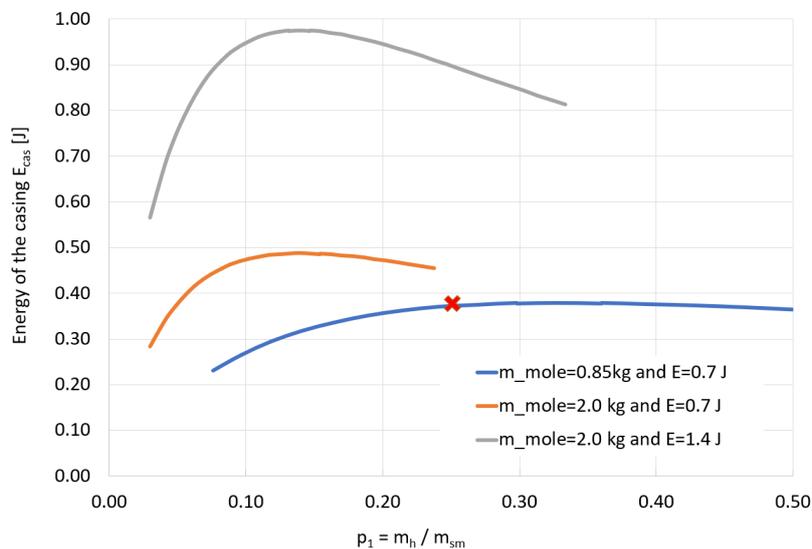

Figure 11. Kinetic energy of the mole as a function of the ratio between the mass of the hammer and of the suppressor-mass p1. A constant value of the mass of the casing of 0.28 kg is assumed. The red cross marks the HP$^3$ mole configuration.

There have been attempts to increase the effective friction coefficient between the soil and the HP$^3$ mole given the fact that the recoil force could not be further reduced. But the design solutions proved to have other disadvantages. For instance, prototypes were built and tested that had barbs on the outer hull (Figure 12) but these were found to increase the effective cross-section area of the mole and thus be disadvantageous for penetration. Once extended, sand filled in behind the barbs, impeding their folding back towards the mole body and increasing the mole's effective diameter. In addition, a risk was seen that the barbs could break off and damage the science tether.

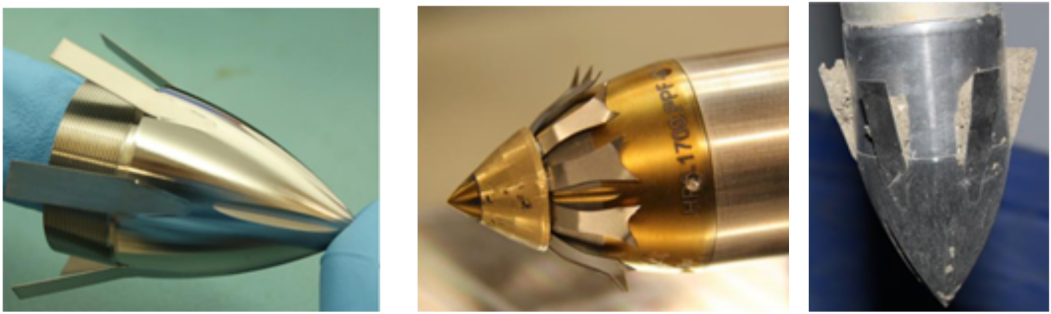

Figure 12. Three prototype mole tips with barbs that were considered to reduce recoil.

Another design solution that has not yet been built or tested would be a mechanism to track the mole into the group and provide rebound resistance to some subsurface depth. The mechanism could consist of a tubular boom, as illustrated in Figure 13, that would add a small but sufficient force to the back-cap of the mole. The tubular boom would be driven by a motor, so that it would be retractable after insertion.

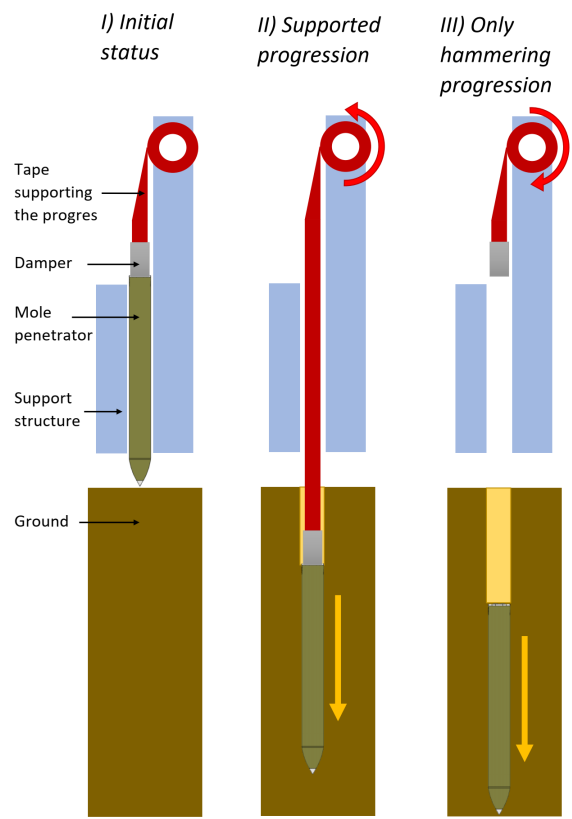

Figure 13. Operational concept of a tubular boom securing full engagement of the mole with the regolith.

A tubular boom, unwound from a reel by a motor, could easily provide at least 10 N of continuous pushing force through an interface damper. The boom should be lightweight but stiff, such as a 0.1 mm thick beryllium bronze tape. The interface damper should be flexible to withstand any recoil and backward impulses generated by the mole. After the mole reaches a suitable depth, the tape-boom and damper could be wound back on the reel. From this point on, the mole would progress without further external support. The science tether with sensors and electrical lines (not shown in Figure 13) would be outside the interface between the damper and the boom. A similar boom has been used as a deployment device for the MUPUS penetrator on the Rosetta lander Philae (Grygorczuk et al. 2007, 2011; Spohn et al., 2009, 2015). For such a subsurface assistance device to function, it would be required that the support structure would either have a sufficiently large mass to resist the transferred rebound forces, or would be pinned to the surface by a manipulator arm or by suitable anchors.

A support mechanism for the mole to balance the recoil comes with a penalty in mass and complexity, of course. Such a solution must be evaluated against the previously discussed alternative of investing mass in the mole hammer mechanism to reduce the recoil which may be more effective.

Other recommendations center around a better diagnostics of mole progress and mole-soil interactions. A soil pressure sensor or strain gauge on the mole's hull and a force transducer at the mole-tether-interface would be useful to get more insight into the actual interplay between the local environment and the penetration performance. Such an equipment would also be useful as scientific instrumentation to measure soil mechanical parameters underneath a planetary surface.

The capability of adjusting the power of the mole may be advantageous in layerred soils. The EMOLE (Grygorzuk et al., 2016) consists of a stack of electromagnets. The EMOLE is driven by the same principle as the previously developed MUPUS (Grygorczuk et al. 2007, 2011; Spohn et al., 2009, 2015) for the Rosetta mission to comet Chyuriumov-Gerasimenko.  The desired level of energy is accumulated in a capacitor and released through electromagnets which accelerate the hammer relative to the suppressor mass. For the EMOLE an energy range of 0–2.2 J, over 3 times more than the energy of the HP$^3$ mole, was achieved but as much as 4 J seem feasible. As a further advantage, the EMOLE system consists of fewer parts arranged in a simpler way than in the spring-driven solution. To be used in the most efficient way, the EMOLE system would need to sense and report knowledge of the device's penetration rate over a small number (e.g., <50) of strokes (Lesson #1),

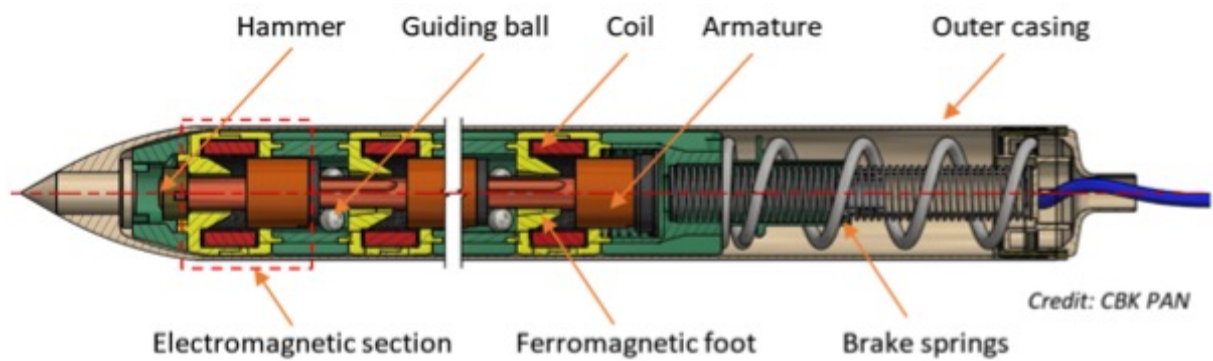

Figure 14 The CBK Warsaw EMOLE system. Shown is a cross section drawing with the stacked electromagnetic hammers.

Ultimately, the HP$^3$ was a bold experiment, attempting to reach unprecedented depth in the Martian regolith with a very compact, low power, and low mass mechanism. The effort proved to have lower performance margins than originally planned (with challenges from the initial penetration, and from the lower atmosphere pressure) and the system encountered an environment more difficult (deep cohesive duricrust) than expected. A more elaborate or massive design may have been able to meet these challenges, but at the expense of more mass and likely at greater cost.  A further dimension in which the effort was challenged was the operations schedule: there was pressure to achieve operations depth ahead of the shadowing thermal wave, and deployment delays eroded the time available. This in turn motivated more aggressive (i.e., longer) hammering commands during the initial penetration sols, which may have been detrimental.  More generous margins in any of these dimensions may have allowed success, but the competed mission framework does not foster large margins.

## References

Baratoux D, H Samuel, C Michaut et al. (2014) Petrological constraints on the density of the Martian crust, J. Geophys. Res. Planets, doi: 10.1002/2014JE004642

Belleguic V, P Lognonne, M Wieczorek (2005) Constraints on the Martian lithosphere from gravity and topography data, J. Geophys. Res. Planets, doi: 10.1029/2005JE002437

Delage, P, F Karakostas, A Dhemaied et al. (2017) An investigation of the mechanical properties of some martian regolith simulants with respect to the surface properties at the InSight mission landing site. Space Science Reviews doi: 10.1007/s11214-017-0339-7

Ehlmann BL, KS Edget, B Sutter et al. (2018) Chemistry, mineralogy, and grain properties at Namib and high dunes, Bagnold dune field, Gale crater, Mars: a synthesis of curiosity rover observations. J. Geophys. Res., Planets, doi:10.1002/2017JE005267

Goetz, W., W. T. Pike, S. F. Hviid et al. (2010) Microscopy analysis of soils at the Phoenix landing site, Mars: Classification of soil particles and description of their optical and magnetic properties. J. Geophys. Res., doi: 10.1029/2009JE003437

Golombek MP, AFC Haldemann, RA Simpson, et al. 2008 Martian surface properties from joint analysis of orbital, Earth-based, and surface observations. *In* J. Bell, ed., The Martian Surface, p 468-498, Cambridge University Press, Cambridge

Golombek MP, D Kipp, N Warner et al. (2017) Selection of the InSight landing site. Space Sci Rev, doi: 10.1007/s11214-016-0321-9.

Golombek MP, M Grott, G Kargl et al. (2018) Geology and physical properties investigations by the InSight Lander. *Space Science Reviews*, doi: 10.1007/s11214-018-0512-7

Golombek, MP, NH Warner, JA Grant et al. (2020) Geology of the InSight landing site on Mars, Nature Communications, 1014, doi: 10.1038/s41467-020-14679-1

Golombek MP, D Kass, N Williams et al. (2020). Assessment of InSight landing site predictions.*Journal of Geophysical Research: Planets*, doi: 10.1029/2020JE006502

Gromov, V, A Misckevich, E Yudkin, H Kochan, P Coste, E Re (1997) The mobile micro penetrometer: a "Mole" for sub-surface soil investigation. Proc. 7th E. Space Mech. Trib. Symp. Edited by B. H. Kaldeich-Schürmann. ESA SP-410. Paris: European Space Agency, 1997., p.151

Grott, M (2009) Thermal disturbances caused by lander shadowing and the measurability of the Martian planetary heat flow. Planet. Space Sci. doi: 10.1016/j.pss.2008.11.005

Grott, M, T Spohn, SE Smrekar et al. (2012) InSight: constraining the Martian heat flow from a single measurement, in 43rd LPSC, The Woodlands, Texas, abstract #1382

Grott M, T Spohn, J Knollenberg et al. (2019) Calibration of the Heat Flow and Physical Properties Package (HP$^3$) for the InSight Mars mission. Earth Space Sci., doi: 10.1029/2019EA000670

Grott, M., T. Spohn, J. Knollenberg et al. (2021) Thermal conductivity of the Martian soil at the InSight landing site from HP$^3$ active heating experiments, J. Geophys. Res. Planets, doi:10.1029/2021JE006861

Grygorczuk J, M Banaskiewicz, K Seweryn, T Spohn (2007). MUPUS Insertion device for the Rosetta mission. *J. Telecom. Inf. Tech.* 1/2007, 50 - 53.

Grygorczuk, J, M Banaskiewicz, A Chichocki et al. (2011). Advanced penetrators and hammering sampling devices for planetary body exploration. In: Proc. of 11th Symposium on Advanced Space Technologies in Robotics and Automation ASTRA 2011. ESA/ESTEC, Noordwijk, 2–14 April 2011, the Netherlands.

Grygorczuk, J., B. Kedziora, L. Wisniewski et al. (2016). A Multi-Sectioning, Reconfigurable Electromagnetic Hammering Propulsion for Mole Penetrators. In: Proc 43rd Aerosp. Mech. Symp. AMS 2016, NASA Ames Research Center, 4-6 May 2016, Santa Clara, CA, USA

Herkenhoff, KE, MP Golombek, EA Guinnes et al. (2008) In situ observations of the physical properties of the Martian surface, *in* J. Bell, ed., The Martian Surface, p 451-467, Cambridge University Press, Cambridge.

Hudson, TL, R Deen, E Marteau, M et al. (2020) InSight HP$^3$ mole near-surface motion and subsurface implications (expanded abstract): 51$^{st}$ Lunar and Planetary Science, Abstract #1217, Lunar and Planetary Institute, Houston.

Klinkmüller, M, G. Schreurs, M. Rosenau, H. Kemnitz (2016). Properties of granular analogue model materials: A community wide survey, Tectonophysics, doi: 10.1016/j.tecto.2016.01.017.

Krömer O, M Scharringhausen, M Fittok et al (2019) Design details of the HP$^3$ mole onboard the InSight mission, Acta Astronautica, doi:10.1016/j.actaastro.2019.06.031

Lorenz, R. D. (2018). Atmospheric test environments for planetary in-situ missions: never quite "Test as you fly", Adv. Space Res., 62, 1884-1894.

Lorenz, R. (2019) Calculating risk and payoff in planetary exploration and life detection missions, Adv. Space Res. doi: 10.1016/j.asr.2019.05.026

Maki JM, M Golombek, R Deen et al. (2018) The color cameras on the InSight lander. Space Sci. Rev., doi: 10.1007/s11214-018-0536-z

Marteau, E, M Golombek, C Vrettos et al. (2021) Soil mechanical properties at the InSight landing site, Mars (expanded abstract): 52$^{nd}$Lunar and Planetary Science, Abstract #2067, Lunar and Planetary Institute, Houston.

McGlynn, I.O., C.M. Fedo, H.Y. McSween Jr., Origin of basaltic soils at Gusev crater, Mars, by aeolian modification of impact-generated sediment. J. Geophys. Res., doi: 10.1029/2010JE003712

Minitti, ME, LC Kah, RA Yingst et al. (2013) MAHLI at the Rocknest sand shadow: science and science-enabling activities. J. Geophys. Res., Planets **118**(11), 2338–2360 (2013)

Moore, HJ, RE Hutton, GD Clow et al. (1987) Physical properties of the surface materials of the Viking landing sites on Mars. U. S. Geol. Surv. Prof. Pap. **1389**, 222pp.

Müller N, S Piqueux, M Lemmon et al. (2021) Near surface properties derived from Phobos transits with HP$^3$ RAD on InSight, Mars. Geophys. Res. Lett., *doi:* 10.1029/2021GL093542

Nagihara, S., P. Ngo, V. Sanigepalli et al. (2020), The Heat Flow Probe for the Commercial Lunar Payload Services Program of NASA, 51st Lunar and Planetary Science Conference, Houston, Lunar and Planetary Institute, p. Abstract #1432.

Nagy B, J. Kovacs, A Igneczi et al. (2020) The termal behavious of ice-bearing ground: The highest cold, dry desert on Earth as an analog for conditions on Mars, at Ojos del Salado, Puna de Atacama-Altiplano Region, Astrobiology, doi:10.1089/ast.2018.2021

Pauer M, D Breuer (2008) Constraints on the maximum crustal density from gravity–topography modeling: Applications to the southern highlands of Mars, Earth Planet. Sci. Lett. doi: 10.1016/j.epsl.2008.09.014.

Peters GH, W Abbey, GH Bearman et al. (2008). Mojave Mars simulant—Characterization of a new geologic Mars analog Icarus, doi: 10.1016/j.icarus.2008.05.004

Piqueux, S, N Müller, M Grott et al. (2021) Soil thermophysical properties near the InSight lander derived from 50 sols of radiometer measurements, J. Geophys. Res., doi: 10.1029/2021JE006859

Pike, T, et al., 2011, Geophys. Res. Lett., 38, L24201.

Plesa, A-C, M Grott, N Tosi et al. (2016). How large are present-day heat flux variations across the surface of Mars? *J. Geophys. Res. Planets,* doi: 10.1002/2016JE005126

Plesa, A-C, S Padovan, N Tosi et al. (2018) The thermal state and interior structure of Mars. Geophys. Res. Lett., doi: 10.1029/2018GL080728

Poganski J, N Kömle, G Kargl et al (2017) Pile driving model to predict the penetration of the InSight/HP$^3$ mole into the Martian soil. *Space Sci. Rev.*, doi:10.1007/s11214-016-0302-z

Rahim A, SN Prasad, KP George (2004) Dynamic cone penetration resistance of soils - theory and evaluations. GeoTrans 2004, doi: 10.1061/40744(154)169

Reershemius S, K Sasaki, M Scharringhausen et al. (2019) Structure development of the HP$^3$ instrument support system for the Mars mission InSight. Acta Astronautica, doi: 10.1016/j.actaastro.2019.06.035

Richter L, P Coste, VV Gromov, H Kochan, R Nadalini, TC Ng, S Pinna, HE Richter, KL Yung (2002) Development and testing of subsurface sampling devices for the Beagle 2 lander. Planet. Space Sci., doi:10.1016/S0032-0633(02)00066-1.

Rummel JD, DW Beatty, MA Jones et al. (2014) A new analysis of Mars "Special Regions": Findings of the second MEPAG Special Regions Analysis Group (SR-SAG2). Astrobiology, doi: 10.1089/ast.2014.1227

Salgado R, JK Mitchell, M Jamiolkowski (1997) Cavity expansion aand penetration resistance in sand, J. Geotech. Geoenv. Eng., doi: 10.1061/(ASCE)1090-0241(1997)123:4(344)

Seweryn K, J Grygorczuk, R Wawrzaszek et al. (2014) Low velocity penetrators (LVP) driven by hammering action – definition of the principle of operation based on numerical models and experimental tests, Acta Astronautica, doi: 10.1016/j.actaastro.2014.03.004

Smrekar SE, P Lognonne, T Spohn et al (2018) Pre-mission InSights on the interior of Mars. Space Sci. Rev., doi: 10.1007/s11214-018-0563-9

Spohn T, K Seiferlin, A Hagermann et al. (2007) MUPUS – a Thermal and Mechanical Properties Probe for the Rosetta Lander Philae, Space Sci. Rev., 128, 339-362, doi: 10.1007/s11214-006-9081-2.

Spohn T, J Knollenberg, AJ Ball et al. (2009) MUPUS – The Philae Thermal Properties Probe. In: Rosetta (R. Schulz, Ed.) pp. 651-668, Springer, Berlin.

Spohn T, J Knollenberg, AJ Ball et al. (2015) Thermal and Mechanical Properties of the Near-Surface Layers of comet 67P/Churyumov-Gerasimenko, Science, 349, 6247, doi: 10.1126/science.aab0464

Spohn M Grott, SE Smrekar et al. (2018) The Heat Flow and Physical Properties Package (HP$^3$) for the InSight mission. Space. Sci. Rev.*,* doi: 10.1007/s11214-018-0531-4

Spohn T, T Hudson, E Marteau et al (2021) The HP$^3$ Penetrator (Mole) on Mars: Soil properties derived from the penetration attempts and related activities. Space Sci Rev, to be submitted.

Trebi-Ollennu A, W Kim, K Ali et al (2018)*.* InSight Mars Lander Robotics Instrument Deployment System. Space Sci. Rev. doi: 10.1007/s11214-018-0520-7

Terzaghi K, RB Peck (1947) Theoretical Soil Mechanics, Wiley, New York

Terzaghi K, RB Peck, G Mesri (1996) Soil Mechanics in Engineering Practice, pp 104-106, 241-247, Wiley & Sons, New York, ISBN: 978-0-471-08658-1

Verruijt, A. (2018) An Introduction to Soil Mechanics, Springer International Publishing.

Vrettos C, (2012) Shear strength investigations for a class of extraterrestrial analogue soils, American Society of Civil Engineers, 2012, DOI: 10.1061/(ASCE)GT.1943-5606.0000619

Warner, NH, M Golombek, J Sweeney et al. (2017). Near surface stratigraphy and regolith production in southwestern Elysium Planitia, Mars: Implications for Hesperian-Amazonian terrains and the InSight lander mission. *Space Science Reviews*, 211(1–4), 147–190. https://doi.org/10.1007/s11214-017-0352-x

Wippermann, T, TL Hudson, T Spohn et al. (2020) Penetration and performance testing of the HP3 Mole for the InSight Mars mission. Planet. Space Sci., doi: 10.1016/j.pss.2019.104780

Zhang N, M Arroyoa, MO Ciantia et al (2019) Standard penetration testing in a virtual calibration chamber, Computers and Geotechnics, doi: 10.1016/j.compgeo.2019.03.021

## Acknowledgement
This paper is InSight Contribution Number 234.

## Appendix: The InSight HP3 Mole on Mars

This appendix provides an abridged version of the actions taken on Mars from the first attempts at mole penetration on Sol 92 (Feb. 28, 2019) through the final penetration attempt on Sol 754 (Jan. 09, 2021). A more extensive version of this "Mole Saga" can be found in Spohn et al. (2021), wherein the sequence of steps taken, their rationale, the evolving state of the mole, and the operational constraints on the HP3 penetration anomaly response effort are fully detailed. Spohn et al. (2021) also contains in-depth investigation and discussion of the scientific lessons learned about the InSight landing site subsurface environment.

| Phase | Sols | Description |
|---|---|---|
| Initial Penetration Attempts | 92 – 94 | Two initial hammerings commanded w/ stop triggers of 4 and 5 hours respectively; STATIL reports significant tilt changes; some Support Structure motion observed via footprints. |
| Diagnostics, Support Structure Lift, Pit Characterization, & First Regolith Interactions | 97 – 257 | Imaging campaigns; short 'diagnostic hammerings'; Support Structure grapple, lift, and replace; IDA scoop 'pushes' and 'chops'. Mole tip depth established as ~33 cm (back cap 7 cm above regolith surface, as measured along the mole body) |
| Pinning Campaign 1 | 291 – 318 | Mole is pinned horizontally and vertically – successful penetration of ~5 cm proves there is no obstructing stone. Back cap progresses to ~2 cm above original regolith surface. |
| Reversal 1 | 322 – 325 | Reconfiguration of the arm to protect the ST removes direct contact with the mole, resulting in insufficient resistance to rebound, a mole reversal event extracts ~18 cm of the mole |
| Pinning Campaign 2 | 329 – 380 | Mole is pinned horizontally and vertically – successful and fast penetration permits recovery from the reversal event to approximately the same depth as at the end of Pinning 1 |
| Reversal 2 | 400 – 407 | Mole is pinned with vertical preload only – another reversal event occurs, extracting ~5 cm of the mole. |

| Regolith Interactions and Back Cap Push Campaign | 414 – 645 | Direct pre-load of the mole back cap. Penetration is successful but slow, resulting in a back cap depth ~2 cm *below* the original regolith surface. Some IDA scoop interaction tests are performed including one large pit-filling scrape. |
| Scrape, Fill, and Final Free Mole Test | 659 – 754 | Scrapes and Tamps bring more regolith into the pit around and above the mole. The scoop is positioned to prevent mole reversal and 500 strokes are commanded. No forward motion is observed; the mole cannot move 'freely' without IDA help. |

Table A.1  Mole Saga phase names, sol intervals, and brief summary of actions and events.

Table A.1 names the major phases of the Mole Saga, the sols covered by each phase, and the major activities and results of the events within the phases. Note to the reader: Unless otherwise specified, mole depths are reported in terms of the distance from the mole back cap to the original regolith surface, as measured along the mole body (i.e., 'along-mole distance' or 'along-mole depth'). Vertical tip depths underground can be determined by subtracting the reported along-mole distances from the total mole length (40 cm) and multiplying the result by the cosine of the mole tilt. IDA: Instrument Deployment Arm; ST: Science Tether

A.1 Initial Attempts: Sols 92 & 94

InSight landed on the surface of Mars on November 18, 2018 and the HP$^3$ package, initially mounted to the lander deck, was deployed to the surface on February 11, 2019 (Sol 76) using the InSight lander's instrument deployment arm IDA (Trebi-Ollenu et al., 2018). The mole was released from its launch lock on Sol 87, committing the instrument to the chosen site. The mole dropped under gravity from its locked position, allowing the tip to penetrate the regolith by ~1 cm.

On Sol 92 the first hammering was commanded with a target depth of 70 cm. Due to the length of the Science Tether service loop between the Tether Length Measurement device TLM and the back cap of the mole (~29 cm), the Science Tether ST was not expected to engage in the TLM until a mole tip depth of ~54 cm below the original regolith surface was achieved. The target depth of 70 cm was chosen to allow absolute depth markings on the science tether to be read by the TLM and provide the stop-hammering command to the HP$^3$ electronics. Data from penetration tests at DLR Bremen and at JPL in a variety of regolith simulants (Wippermann, 2020) led to the expectation that the mole would reach this target depth within 30 minutes of hammering. A nominal hammering timeout period of 4 hours was set to allow for the possibility that the shallow regolith was more difficult to penetrate than in any of the terrestrial tests.

The penetration anomaly was recognized when data indicated that the mole hammered for the full 4 hours (3881 hammer strokes) and no data was reported from the TLM. Other available data included: (1) The mole back cap passed the contact switch (24.5 cm above regolith surface = mole tip depth of 15.5 cm) 4 minutes 54 seconds (77 strokes) after hammering began. (2) STATIL reported significant tilt changes, with the greatest magnitude occurring in the first 11 minutes (~170 strokes) (See Figure XX). (3) At an unknown time or times during the interval, the SS moved along its y-axis towards the InSight lander by ~1.75 cm, as revealed by footprint markings of the initial placement site.

Informed by terrestrial penetration experiments, the team surmised there might be over-large resistance in the regolith (either a stone or unexpectedly dense material) was a likely culprit for the poor penetration performance. So, on Sol 94 a second hammering period was commanded with the same 70 cm target tip depth and a 5-hour timeout. As with the first attempt, no readings were recorded by the TLM, and the mole timed out after 5 hours (4720 strokes).

During the second interval, the following were observed: (1) The overall tilt as reported by STATIL fluctuated around an average of 18 degrees, with minor and near-instantaneous noise excursions and discontinuities in all sensors. (2) At an unknown time or times during the second interval, the SS again was moved, leaving footprint impressions in the regolith. This time the movement was a rotation of ~4 degrees anti-clockwise around a pivot at the rear edge of the aft right foot.

A.2 Diagnostics, Support Structure Lift, Pit Characterization, & First Regolith Interactions (97 – 257)

During the 160 sols from Sol 97 to 257, numerous activities were conducted to understand the mole anomaly and take action to assist the mole in penetration. Prior to the removal of the Support Structure and exposure of the mole, a multitude of images were acquired with the lander-fixed Instrument Context Camera ICC and the arm-mounted Instrument Deployment Camera IDC (see Maki et al., 2018 for a description of both cameras). Some short (~200 strokes) diagnostic hammerings were also performed before the Support Structure was lifted. These hammerings did not significantly alter the tilt of the mole and there is no evidence that further penetration was achieved. The SEIS instrument adopted a digital filter configuration to more precisely listen to the character of the diagnostic hammering strokes as a further diagnostic tool (see Spohn, et al. 2021).

Though the risk of a Science Tether snag within the Support Structure could not be ruled out, the project decided to proceed with re-grappling and removing the structure to expose the mole for further imaging and interaction with the robotic arm. The lift was successful, extracting more ST and clearly revealing the mole azimuth, depth, and the surprisingly steep-walled pit (Figure 3).

All the various diagnostic evidence collected during this period pointed to a healthy mole, functioning as designed and with the expected stroke energy. The only credible root-causes remaining to explain the mole's lack of penetration were (1) an External Obstruction, i.e., the mole was obstructed by a dense regolith layer, pathologically sized gravel, or a large stone; or (2) Lack of Sufficient Friction between the mole hull and the regolith. Or perhaps both.

Some regolith interactions were attempted with the IDA scoop to collapse regolith material into the pit to increase friction on the hull, but the cohesive layer proved resistant to these initial 'light touch' attempts. This prompted the first of two 'pinning' campaigns.

A.3 Pinning 1 (291-318)

The 'pinning' activities were motivated by the desire to increase friction on the mole's hull (in truth the only option, since nothing could be done about a subsurface obstruction). Both pinning campaigns used the edge of the IDA scoop to pre-load the side-hull of the mole with the maximum achievable safe force, providing friction at the IDA point of contact, and also by pushing the mole into the regolith opposite the push-point. The amount of force applied to the mole is no greater than the maximum measured using the flight-spare arm in the InSight deployment testbed at JPL: at most ~40 N in the vertical direction, and at most ~25 N in the horizontal direction. It was hoped that this increased friction would exceed the 5 – 7 N rebound force threshold and allow the mole to make forward progress… again, only if there were no obstructing stone.

After positioning the arm and pre-loading the mole both vertically and horizontally, the first hammer strokes in 150 sols (since Sol 158) were commanded on Sol 308. The resulting IDC movie showed some barely perceptible (~5 mm) motion downward and rotation around the mole's long axis. There was no significant change in overall tilt reported by the STATIL sensors, though they did confirm that the slight rotation of the mole during these 20 strokes was a real effect. The tendency of the mole to rotate during hammering had been observed in the laboratory.

On Sol 311, 315, and 318, further hammerings of 101, 101, and 152 strokes were commanded, resulting in a combined mole penetration of ~ 5 cm as measured along the mole (see Spohn, et al. 2021 for a description of the method used to extract mole depth from IDC images). This significant progress, accompanied by only minor changes in mole tilt, conclusively ruled out an impassable external obstruction (e.g., a large stone). Recall that the IDA was not pushing the mole into the regolith but only loading it from the side; all downward progress was due to mole hammering alone.

A.4 Reversals (322 - 325)

Unfortunately, tactics now needed to change as the mole back cap approached the regolith surface. To avoid potential damage to the ST from a sideswipe by the scoop, the direct pre-load of the mole was removed and the underside of the scoop was pushed flat against the adjacent regolith. Analysis suggested that some IDA force could be transferred to the mole via a regolith-mediated load path thereby increasing friction on the mole below the bottom of the pit (~30 cm).

These analyses proved to be incorrect when, on Sol 322 and 325, a total of 354 hammer strokes were commanded but resulted not in further penetration or even 'bouncing' in place, but a rapid and near-catastrophic self-extraction of the mole from the regolith. The mole backed out of the ground by ~18 cm. The mole tilt (which had remained in the range of 19±1 degrees since the end of Sol 92) changed very little over most of the extraction, but then experienced a sharp increase of nearly 7 degrees (reaching 24 degrees from vertical) in the final ~100 strokes.

A.5 Pinning 2

The mole penetration anomaly response team reacted rapidly to the reversal event, converging on a solution to re-attempt the successful pinning method by Sol 346. Five further hammering sols (346, 349, 366, 373, 380; with 40, 50, 19, 127, 126 strokes respectively) with a horizontally & vertically pre-loaded mole hull resulted in rapid re-penetration of the mole, from an extreme back-cap height of ~18 cm to ~3 cm (nearly the same depth as at the conclusion of Pinning 1) by Sol 380. The maximum rate of motion during this re-penetration (compare Fig. A1) occurred over Sols 349 and 366 where it was computed to be ~0.6 mm/stroke. The hammerings on Sols 373 and 380 had an average re-penetration rate of 0.3 mm/stroke. It should be noted though that the re-penetration during Pinning 2 occurred with a mole tilt of 24 – 27 degrees from vertical, whereas Pinning 1 was conducted with the mole tilt in the range of 18 – 20 degrees.

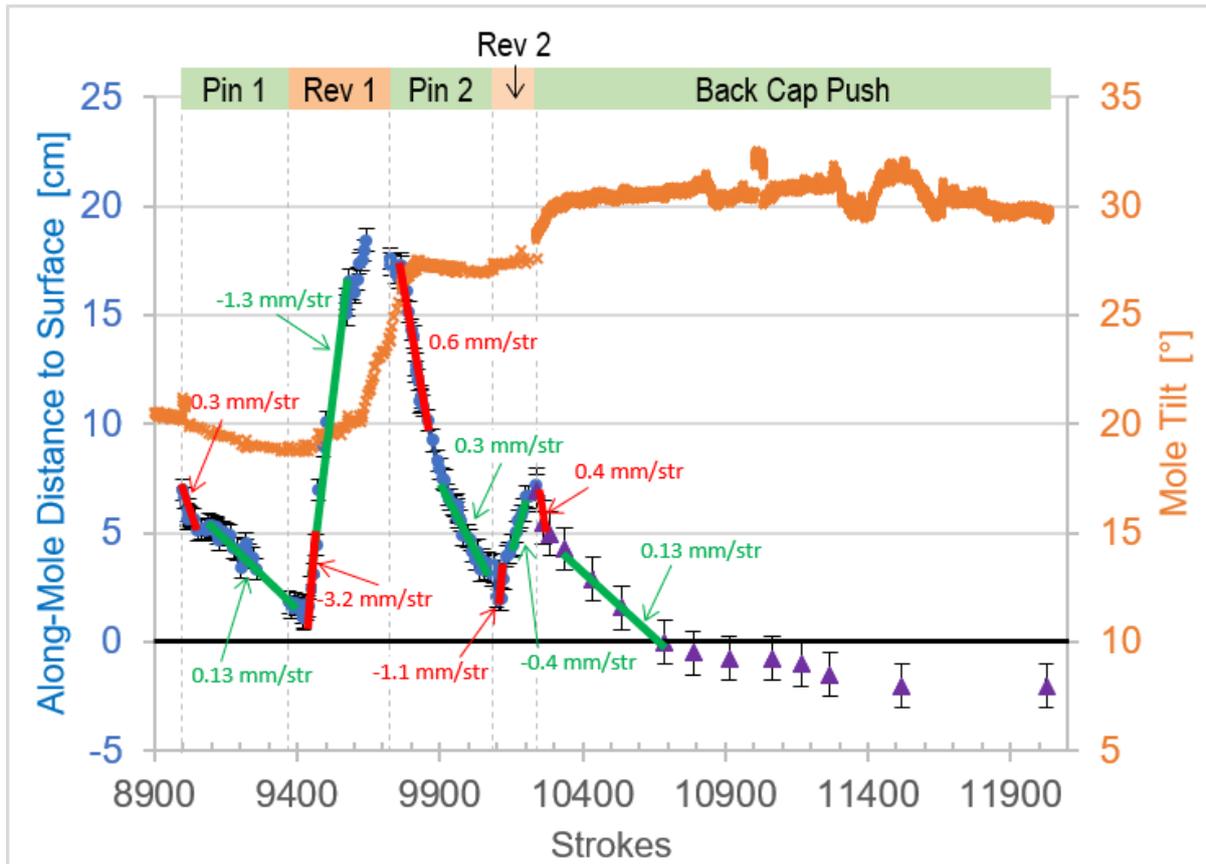

Figure A1. History of the mole down- and upward movements during the efforts of trying to get the mole to bury to depth and of the mole inclination. The phases of pinning, reversals and back cap push are marked and discussed in the text.

Now the team faced the same problem as at the end of Pinning 1: potential swipe-damage to the Science Tether. Instead of attempting a regolith pre-load, the arm was commanded to do a vertical-only retraction and re-application of pre-load to the tilted mole hull. This new method also proved unsuccessful when the 151 strokes commanded on Sol 407 resulted in 5 cm of along-mole extraction at approximately constant tilt.

A.6 Regolith Interactions and Back Cap Push (414 - 645)

Rather than re-attempt side-pinning and encounter the same problem again, the team shifted tactics to a more challenging paradigm that offered lower self-extraction risk. The contextually low precision of the InSight IDA, the small size and irregular shape of the mole's back cap, the presence of the delicate-in-shear science tether, and the decreasing resources available to the project had together deferred the direct application of along-axis pre-load at the mole's aft end until this point. But having no other options, the team proceeded with the long Back Cap Push campaign (fully detailed in Spohn, et al. 2021).

The Back Cap paradigm had the advantage of direct pre-load supplanting friction as the main source of resistance to mole rebound. It also placed the scoop in the path of the mole's rebound vector, directly mitigating the risk of reversal. This delicate operation required much finer positioning than was typical for the arm and each placement was approached carefully to do no harm to the science tether, mole, or IDA. Since the geometry of the mole penetration vector and the robotic arm actuators prevented the scoop from directly following the mole along its path into the regolith, each hammering period was followed by a repositioning of the scoop and a re-application of IDA preload.

From the time of Reversal 2 (Jan 11, 2020) it took a little over 8 months to conduct 12 back cap hammerings (9 with a horizontally oriented scoop, 3 with an inclined scoop), totaling 1280 hammer strokes. That effort brought the mole back cap from its position at the end of Reversal 2 (7 cm along-mole distance to the original regolith surface) to a back cap depth of ~ 2 cm *below* the original regolith surface on Sol 645 (Sep 19, 2020). Having insufficient resources to widen the pit, this was the maximum depth that could be reached by the IDA scoop. It should be noted that in some of the Back Cap push hammerings (horizontal scoop Sols 550 and 557; inclined scoop Sols 632 and 645) no apparent downward motion of the mole (as proxied by the above-ground science tether) was observed. During these hammerings, it was observed that small regolith particles in the scoop would be static during the initial hammering, but then would begin 'jumping' around from image to image. This is indirect evidence that the mole was attempting to hammer 'free' from the scoop but did not have sufficient friction to progress on its own and ended up bouncing into the scoop (and would have self-extracted had the scoop not been in place).

Interleaved with the Back Cap campaign's preloaded hammerings were several further direct interactions of the scoop and regolith. Most of these were engineering tests in anticipation of what would be needed when the arm could no longer follow the mole into the subsurface. The most impactful of these occurring before Sol 645 was a 'chop' on Sol 420 where a small piece of consolidated subsurface material was broken off and fell into the pit. and a scrape on Sol 598 which brought enough material into the pit to nearly obscure the mole.

A.7 Scrape, Fill and Final Free Mole Test (659 - 754)

Following the last of the Back Cap push efforts on Sol 645, a further series of 'scrapes' (Sols 673 & 700) and 'tamps' (Sols 686 & 734) brought unconsolidated surface material into the pit and compacted it. The scrapes and tamps were performed to maximize the amount of regolith in contact with the mole, increasing regolith friction so the mole could make downward progress without assistance from the robotic arm.

By the time these were completed (Sol 734, Dec. 19, 2020), power and thermal considerations for InSight were complicating operations as dust continued to accumulate on the solar panels and Mars approached aphelion. The history of the mole's low penetration rates (typically 0.1 mm/stroke) had to be considered in the context of the expected lifetime of InSight. In the best recovery scenario envisioned at the time, the mole would need to dig uninterrupted until it could no longer dig or reached its minimum required depth of 3 meters. Heat from hammering would dissipate during the aphelion lull and conjunction. Then, assuming InSight survived, some clean measurements of the thermal gradient could be made when operations resumed.

Working backwards from these and other constraints, it was decided that there would be one final test, a "Free Mole Test", after the scrapes and tamps of the previous period were completed. In this test, the IDA was maximally preloaded onto the regolith above the mole which was then be commanded to hammer 500 strokes, the high number being chosen such that the result, whatever it was, would be unambiguous. The test occurred on Jan 9, 2021 (Sol 754). No further downward motion was detected, the mole tilt varied irregularly between 32° to 29.5°, and regolith particles on the IDA and in the scoop were seen to move erratically. This latter evidence suggests the mole was attempting to reverse and rebounding into the scoop, similar to what was seen during some back cap push activities. Thus, it was determined that the final Free Mole Test was not successful and further attempts to assist the mole achieve greater depth were abandoned.